\documentclass[smallextended]{svjour3}  
\usepackage{graphicx}                
\usepackage{amsmath, amssymb}        
\usepackage[numbers]{natbib}                 
\usepackage{pdflscape}
 \usepackage{longtable}
 \usepackage{hyperref} 
 \usepackage{lscape} 
 \usepackage[table,xcdraw]{xcolor}
 \usepackage[normalem]{ulem}
 \useunder{\uline}{\ul}{}
\usepackage{rotating} 
\usepackage{lmodern} 
\usepackage{booktabs}

\begin{document}

\title{Empirical Evaluation of AI-Assisted Software Package Selection: A Knowledge Graph Approach}
\titlerunning{PySelect}  

\author{Siamak Farshidi \and Amir~Saberhabibi \and Behbod Eskafi \and Niloofar Nikfarjam \and Sadegh Eskandari \and Slinger Jansen \and Michel Chaudron \and Bedir Tekinerdogan}
\authorrunning{Farshidi et al.} 
\institute{
Siamak Farshidi, Bedir Tekinerdogan \at Information Technology Group, Wageningen University \& Research, Wageningen, The Netherlands \\
\email{\{siamak.farshidi,bedir.tekinerdogan\}@wur.nl}
  \\
Amir Saberhabibi, Behbod Eskafi, Niloofar Nikfarjam, Sadegh Eskandari \at Department of Computer Science, Guilan University, Rasht, Iran \\
\email{eskandari@guilan.ac.ir}
\\
Slinger Jansen \at Department of Information and Computer Science, Utrecht University, Utrecht, The Netherlands\\
\email{slinger.jansen@uu.nl}
\\
Michel Chaudron \at Department of Mathematics and Computer Science,  TU Eindhoven, Eindhoven, The Netherlands\\
  \email{m.r.v.chaudron@tue.nl}
}

\date{Received: date / Accepted: date}

\maketitle

\begin{abstract}

\textit{Context:} Selecting third-party software packages in open-source ecosystems such as Python is challenging due to the significant number of alternatives and the limited availability of transparent evidence to compare them. Generative AI tools are commonly used in development workflows, yet their suggestions often overlook dependency evaluation, prioritize popularity over suitability, and lack reproducibility. This presents risks for projects that require transparency in decision-making, long-term reliability, maintainability, and informed architectural decisions.

\textit{Method:} This study formulates software package selection as a Multi-Criteria Decision-Making problem and proposes a data-driven framework for technology evaluation. Based on the framework, automated data pipelines are designed to continuously collect and integrate software metadata, usage trends, vulnerability information, and developer sentiment from GitHub, PyPI, and Stack Overflow. These data are structured into a decision model that represents relationships among packages, domain features, and quality attributes. The framework is implemented in \textit{PySelect}, a decision support system that interprets user intent using large language models and queries the decision model to identify contextually appropriate software packages.

\textit{Results:} The proposed approach is evaluated using 798{,}669 Python scripts from 16{,}887 GitHub repositories, along with a user study based on the Technology Acceptance Model. The results show high precision in data extraction, improved recommendation quality compared to generative AI baselines, and positive user evaluations of perceived usefulness and ease of use.

\textit{Contribution:} This work introduces a scalable, interpretable, and reproducible framework that supports evidence-based software package selection through Multi-Criteria Decision-Making principles, empirical data, and AI-assisted intent modeling. By addressing key limitations in current generative AI workflows, the proposed approach enables more reliable, reproducible, and quality-aware software engineering decisions.

\keywords{Software Package Selection \and Multi-Criteria Decision Making \and Knowledge Graph \and decision-support systems \and Python Ecosystem \and Empirical Software Engineering \and Generative AI}

\end{abstract}

\section{Introduction}

The rapid expansion of open-source ecosystems has fundamentally reshaped software engineering, enabling developers to build upon shared tools and collective knowledge at an unprecedented scale~\cite{wang2023power}. In ecosystems such as Python, developers have access to a vast collection of reusable third-party software packages, which support innovation by reducing redundancy and leveraging community-driven solutions~\cite{Zajdel2022OSS}. These packages are distributable units of code that encapsulate functionality and are typically shared through centralized package managers such as the Python Package Index (PyPI), which provides metadata, versioning, and installation support~\cite{bavota2015apache,rahkema2024impact,islam2023empirical}. PyPI alone hosts more than 660{,}000 distinct projects, over 7 million releases, and nearly 15 million distribution files, and receives more than 13 billion downloads per month~\cite{pypistats2025}. This immense scale highlights the richness of available alternatives but also complicates the task of identifying the most appropriate package for a given context. Software packages differ substantially in quality, documentation, community support, security posture, and maintenance activity~\cite{Zajdel2022OSS,akhavani2025open,jadhav2009evaluating}. This variation introduces significant risks, particularly in large-scale or safety-critical contexts, where system dependability hinges on consistent and well-maintained components~\cite{fritz2024vulnerability,berntsson2017evaluation}.

As open-source ecosystems continue to expand, generative AI tools such as GitHub Copilot~\cite{githubcopilot} and ChatGPT~\cite{openai2023chatgpt,openai2023gpt4} have become influential components of modern software development workflows. These models translate natural language prompts into executable code, assist with syntax, and suggest implementation patterns, leading to notable improvements in developer productivity, particularly for routine programming tasks~\cite{pandey2024transforming}. Despite these benefits, such tools often lack contextual awareness. They typically assume that appropriate dependencies have already been selected, without assessing whether those choices satisfy requirements for quality, performance, or long-term maintainability~\cite{li2024assessing}. In addition, generative models operate as opaque systems: their internal reasoning processes are not transparent, their outputs may hallucinate non-existent software packages or APIs, and their recommendations often reflect biases toward widely used or well-documented packages rather than those best suited to the task~\cite{pearce2025asleep,spracklen2024we}.

These limitations raise critical concerns when AI-generated recommendations influence dependency choices during the development process. Developers may inadvertently adopt suboptimal or vulnerable software packages because the recommendations lack transparency, accountability, or validation against real-world dependency data~\cite{choudhuri2025needs,sabouri2025trust}. Such practices can compromise architectural integrity, introduce security vulnerabilities, and erode trust in generative tools as reliable guides for making critical technical decisions~\cite{afroogh2024trust}. Consequently, there is a pressing need for a transparent, reproducible, and evidence-based framework for selecting software packages.

To address the limitations of current software package recommendation approaches, this study proposes a decision-support framework grounded in multi-criteria decision-making (MCDM) and empirical software engineering. The framework integrates metadata from diverse sources, including code repositories, package registries, vulnerability databases, and developer forums. Automated pipelines extract and normalize usage patterns, quality indicators, and semantic factors, which are then mapped to standardized quality models and embedded within a structured knowledge graph.

The framework is instantiated in a software artifact named \textit{PySelect}, which operationalizes the proposed approach and provides an interactive interface for technology selection. PySelect enables users to explore package alternatives based on empirical criteria such as maintenance activity, adoption trends, known vulnerabilities, and community sentiment. The artifact serves as both a decision-support tool and a vehicle for evaluating the framework’s effectiveness in practice.

The system is evaluated through large-scale analysis of 798{,}669 Python scripts across 16{,}887 GitHub repositories and a user study involving 22 participants. Its recommendation quality is further assessed through comparative analysis against established generative AI models, including ChatGPT, Copilot, and DeepSeek. These evaluations demonstrate the viability of combining structured data, quality models, and intent modeling to support transparent and context-sensitive software package selection.

\vspace{1em}
\noindent \textbf{The main contributions of this study are as follows:}

\begin{itemize}
    \item A theoretical framework for software package evaluation that integrates MCDM with large-scale empirical evidence and software quality standards.
    
    \item PySelect, an operational artifact that realizes the framework through automated data pipelines, a software knowledge graph, and a user-facing recommendation interface.
    
    \item A cross-source extraction method that consolidates metadata, usage patterns, vulnerability records, and developer sentiment into a unified evaluation model.
    
    \item A curated knowledge graph encompassing 39{,}841 unique Python packages, annotated with contextual, semantic, and quality-related features.
    
    \item An empirical evaluation, including a comparative study with generative AI tools and a user study based on the Technology Acceptance Model, demonstrating the system’s effectiveness and usability.
\end{itemize}

The remainder of this paper is organized as follows. Section~\ref{sec:ResearchApproach} defines the problem, formulates the guiding research questions, and outlines the design science methodology that frames the study. Section~\ref{sec:Framework} introduces PySelect, detailing its system architecture, data extraction pipelines, knowledge graph construction, and recommendation mechanism. Section~\ref{sec:evaluation} presents a multi-dimensional evaluation of PySelect, including precision and recall metrics for its data pipelines, a comparative analysis against leading generative AI tools (ChatGPT, Copilot, and DeepSeek), and a user study based on the Technology Acceptance Model (TAM). This section also includes an empirical characterization of 798{,}669 Python scripts contained in the PySelect knowledge base, which was constructed by mining open-source software repositories. The analysis highlights long-tail usage patterns, domain keyword distributions, and the availability of packages on and off PyPI. Section~\ref{sec:dicussion} synthesizes the findings, addresses the research questions, and reflects on broader implications, limitations, and lessons learned. Section~\ref{sec:RelatedWork} situates this work within the broader literature on software package selection, quality modeling, and developer support tools. Finally, Section~\ref{sec:conclusion} summarizes the key contributions and outlines future directions, including support for other ecosystems and more refined modeling of dependency relationships.

\section{Research Approach}~\label{sec:ResearchApproach}

This study adopts a structured research approach to address the challenge of third-party software package selection in large-scale, open-source ecosystems, with a focus on the Python ecosystem. The work integrates artifact-oriented engineering with empirical inquiry, following the design science research methodology~\cite{hevner2004design,peffers2007design}. The objective is to design, implement, and evaluate a theoretical MCDM-based framework that models data collection, integration, and recommendation in this context. To operationalize this framework, we develop \textit{PySelect}, a data-driven decision-support system that assists software practitioners in selecting packages aligned with project-specific requirements.

The research approach is guided by a clearly defined problem formulation and three targeted research questions. These questions are addressed through an iterative process that involves: (1) designing automated data collection and integration pipelines, (2) constructing a structured knowledge graph that forms the basis of a decision model, and (3) evaluating the effectiveness and usability of the system. Supporting methods include a semi-structured literature study to inform design decisions, experimental evaluations for pipeline evaluation and model benchmarking, and a user study grounded in the Technology Acceptance Model~\cite{davis1989perceived}.

The following subsections outline the problem formulation, research questions, and research methods applied to conduct the study.

\subsection{Problem Formulation}~\label{problemFormulation}

Technology selection in software engineering~\cite{pressman2001software,farshidi2018multiple} is the process of identifying tools, components, or services that best satisfy a combination of functional~\cite{wiegers2013software} and non-functional requirements~\cite{chung2012non}. This process occurs across a range of domains, such as architectural patterns~\cite{schmidt2013pattern,ArchitecturePatternsfarshidi2020capturing}, database systems~\cite{Databasefarshidi2018decision}, cloud platforms~\cite{Cloudfarshidi2018decision}, and third-party software packages~\cite{pace2023javascript}. The growing number of available alternatives, coupled with substantial variation in quality, maturity, and ecosystem compatibility, contributes to increased complexity and risk in the selection process.

This problem can be systematically addressed using MCDM theory, which provides a basis for evaluating alternative technologies against structured sets of evaluation criteria~\cite{thakkar2021multi}. In our earlier studies, we employed MCDM theory to develop a theoretical framework for modeling the technology selection problem in software engineering~\cite{farshidi2020multi}. This framework has been instantiated in various contexts. These instances include database management~\cite{Databasefarshidi2018decision}, cloud service provisioning~\cite{Cloudfarshidi2018decision}, blockchain platforms~\cite{Blockchainfarshidi2020decision}, and software architecture patterns~\cite{ArchitecturePatternsfarshidi2020capturing,farshidi2020decision}. Additional applications have extended the framework to model-driven development~\cite{farshidi2021model}, programming language ecosystems~\cite{farshidi2021decision}, decentralized autonomous organizations~\cite{DAObaninemeh2023decision}, business process modeling languages~\cite{BPMLfarshidi2024business}, user intent modeling~\cite{IntentModelfarshidi2024understanding}, and blockchain oracle platforms~\cite{Oracleahmadjee2025decision}.

A key limitation of these earlier instantiations is their dependence on manual data collection, typically through expert interviews and literature reviews. While feasible in domains with a limited and stable set of alternatives, this approach does not scale to rapidly evolving environments such as software package selection. In such contexts, the number of available packages and their characteristics change frequently, and relevant information is distributed across heterogeneous sources. Despite these challenges, the selection process remains well-structured and aligns with the MCDM paradigm, consisting of a defined set of alternatives, domain-specific features, and a mapping between project requirements and solution candidates.

\begin{figure*}[ht]
    \centering
    \includegraphics[trim=22 20 20 20,width=0.95\textwidth]{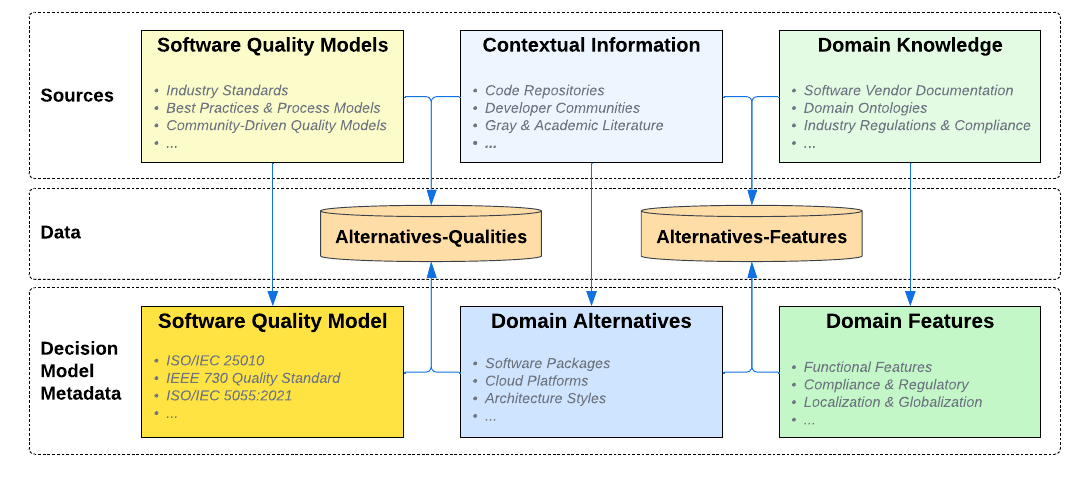}
    \caption{An extension of the technology selection framework~\cite{farshidi2018multiple}. The proposed framework incorporates a data collection perspective that enables automated integration pipelines by linking data sources (such as quality models, contextual information, and domain knowledge) to decision model metadata through structured mappings of alternatives to qualities and features.}
    \label{fig:MCDMinSE}
\end{figure*}

Figure~\ref{fig:MCDMinSE} illustrates the proposed framework, which extends our prior MCDM-based approach~\cite{farshidi2018multiple} by introducing automated data integration mechanisms. The conceptual model is organized into three layers: sources, data, and decision model metadata. The sources layer includes software quality models (e.g., ISO/IEC 25010, IEEE 730), contextual information (e.g., code repositories and developer communities), and domain knowledge (e.g., documentation and standards). These inputs support two key mappings: from alternatives to quality attributes, and from alternatives to domain-specific features.

The decision model metadata layer defines the MCDM structure, encompassing quality standards, domain-specific alternatives (such as software packages or cloud services), and contextual features (such as compliance or integration requirements). This framework enables reproducible, scalable, and context-aware technology selection. By incorporating automated data pipelines, it addresses the scalability limitations of previous manually-driven approaches while maintaining alignment with established evaluation principles.

Software package selection represents a particularly dynamic and complex instance of the technology selection process, especially in open-source ecosystems such as Python. Developers must choose from a large and continuously evolving pool of packages~\cite{pypistats2025}, which vary in reliability, maintenance activity, vulnerability exposure, documentation quality, and adoption trends. This variability, combined with the distributed nature of relevant data, renders manual evaluation infeasible at scale.

To address this complexity, the proposed framework in Figure~\ref{fig:MCDMinSE} is instantiated to support automated and evidence-based evaluation of software packages. In this context, each package is modeled as an alternative described by a vector of measurable features, and the decision process involves identifying those alternatives that align most closely with project-specific quality and functional requirements. Formally, let $\mathit{Packages} = \{p_1, p_2, \ldots, p_n\}$ denote the set of candidate packages, where each package is characterized by a set of attributes $\mathit{Features} = \{f_1, f_2, \ldots, f_m\}$. Project requirements are represented as a subset $\mathit{Requirements} \subseteq \mathit{Features}$. The goal is to identify a solution set $\mathit{Solutions} \subseteq \mathit{Packages}$ that maximizes alignment with the specified requirements. This relationship is captured by the mapping $\text{MCDM} : \mathit{Packages} \times \mathit{Features} \times \mathit{Requirements} \rightarrow \mathit{Solutions}$.

\subsection{Research Questions}~\label{RQs}

Selecting appropriate software packages is a recurring and often difficult task for developers working in large, open-source ecosystems. In the Python ecosystem, where thousands of third-party software packages are available, options differ significantly in terms of quality, maintenance activity, usage trends, and documentation. Developers frequently rely on informal practices such as personal experience, peer suggestions, or simple popularity metrics like GitHub stars and PyPI downloads. These factors, while convenient, may not accurately reflect how well a software package aligns with a project's specific requirements.

To address this gap, there is a need for systematic, data-driven support that helps developers make informed software package selection decisions based on real-world usage data, quality models, and project context.

\textbf{The main research question (MRQ) guiding this study is:}  
\textit{How can software practitioners be supported in selecting appropriate third-party software packages in the Python ecosystem?
}

To answer this question, the study is structured around the following three research questions:

\begin{itemize}
    \item[] \textbf{RQ1:} How can relevant data about software packages in the Python ecosystem be collected automatically, and which sources provide the most useful information on quality, usage, and context?
    \item[] \textbf{RQ2:} How can the collected data be integrated into a structured knowledge graph that forms the basis of a decision model for software package selection?
    \item[] \textbf{RQ3:} How can the resulting decision model be evaluated to determine its effectiveness in supporting developers during software package selection?
\end{itemize}

\subsection{Research Methods}~\label{ResearchMethods}

This study follows a multi-phase methodology grounded in the \textit{Design Science Research (DSR)} paradigm, based on the framework proposed by Hevner et al.~\cite{hevner2004design} and extended by Peffers et al.~\cite{peffers2007design}. The overarching goal is to design, implement, and evaluate a decision-support framework for evidence-based software package selection in the Python ecosystem. This framework is operationalized through the development of a software artifact called \textit{PySelect}.

 designing and reporting this study, we also considered the empirical research standards developed by the ACM SIGSOFT community, specifically the {Empirical Standards for Software Engineering (EMSE)}~\cite{ralph2020empirical}. These standards promote rigor and transparency in empirical software engineering. Where applicable, we adhered to relevant EMSE guidelines for design science studies, controlled experiments, and technology evaluations. This includes providing detailed descriptions of research questions, study context, design rationale, participant selection, data analysis techniques, and threats to validity (see Sections~\ref{sec:RelatedWork},~\ref{sec:data-extraction},~\ref{sec:evaluation}, and~\ref{sec:ThreatstoValidity}).

The research activities are aligned with the three guiding research questions (RQ1–RQ3), as detailed below.

\paragraph{Design Science Process:}

The DSR methodology consists of the following key components~\cite{hevner2004design,peffers2007design}:

\begin{itemize}
    \item \textbf{Problem Identification and Motivation (RQ1):} The research problem is defined based on the absence of scalable and systematic solutions for software package selection. The need for automated data collection emerges from the rapid growth and dynamic nature of metadata in open-source ecosystems like Python.

    \item \textbf{Objective Definition (RQ1 and RQ2):} The objective is to develop a system that collects, integrates, and structures data from heterogeneous sources,including package registries, software repositories, and quality standards, into a knowledge graph that supports decision-making.

    \item \textbf{Artifact Design and Development (RQ2):} The design and implementation of \textit{PySelect} includes automated data pipelines, a structured knowledge graph representing alternatives and features, and an inference engine that aligns packages with project-specific requirements.

    \item \textbf{Demonstration (All RQs):} The artifact is applied in realistic software selection scenarios using real-world project data and user inputs.

    \item \textbf{Evaluation (RQ1–RQ3):} The system is evaluated through experimental validation of the data pipelines (RQ1), comparative analysis of the inference mechanism (RQ2), and a user study assessing practical utility and acceptance (RQ3).

    \item \textbf{Communication:} The research design, artifact, and empirical results are documented in this paper and disseminated to both academic and practitioner communities.
\end{itemize}

\paragraph{Literature Study to Support Design Decisions (RQ1 and RQ2):}

To guide the initial design decisions, we conducted a semi-structured literature study following the snowballing methodology proposed by Wohlin~\cite{wohlin2014guidelines}. The review focused on software quality evaluation, dependency management practices, and decision-support systems in software engineering. Insights from the review informed the design of the metadata schema, feature extraction strategies, and the structure of the decision model, and the details of this study are presented in Section~\ref{sec:RelatedWork}.

\paragraph{Evaluation of Data Collection Pipelines (RQ1):}

An experimental study was conducted to evaluate the reliability and completeness of the automated data collection and metadata extraction processes, following best practices for empirical software engineering experiments as outlined by Wohlin et al.~\cite{wohlin2012experimentation}. Key validation tasks included:

\begin{itemize}
    \item Extraction and resolution of import statements in Python scripts across a diverse codebase.
    \item Mapping to valid PyPI entries or identification as unavailable packages.
    \item Normalization of key metadata fields such as licensing, update frequency, and vulnerability data.
\end{itemize}

\paragraph{Evaluation of Data Integration and Inference Engine (RQ2):}

To evaluate the quality of the knowledge graph and its inference capabilities, we performed a comparative experiment. PySelect's package recommendations were benchmarked against those produced by generative AI models (ChatGPT, GitHub Copilot, and DeepSeek). This evaluation focused on relevance, completeness, and alignment with declared project requirements, using expert-annotated test cases.

\paragraph{User Study Based on Technology Acceptance Model (RQ3):}

To evaluate the perceived usefulness, usability, and acceptance of \textit{PySelect}, we conducted a user study based on the \textbf{TAM} as defined by Davis~\cite{davis1989perceived}. Participants, including software developers and graduate students, performed guided software package selection tasks using PySelect and then completed a structured TAM questionnaire.

The instrument consisted of 15 items organized into four established TAM constructs:

\begin{itemize}
    \item \textbf{Perceived Usefulness}: the extent to which users believe that PySelect helps them make better software package decisions. Example items include: “I believe this platform would help me find better Python packages” and “The platform supports better decision-making when choosing third-party tools.”

    \item \textbf{Perceived Ease of Use}: the degree to which users find the system intuitive and easy to interact with. Example items include: “It was easy to understand how to use the platform” and “It didn’t take much effort to find relevant packages.”

    \item \textbf{Attitude Toward Using}: the user’s general evaluation of the platform. Example items include: “I have a generally positive opinion about the platform” and “I enjoyed testing or exploring this platform.”

    \item \textbf{Behavioral Intention to Use}: the user’s willingness to adopt or recommend the platform in practice. Example items include: “If the platform were publicly available, I would consider using it” and “I would be interested in trying future versions of the platform.”
\end{itemize}

Responses were recorded using a 7-point Likert scale ranging from \textit{Strongly Agree (1)} to \textit{Strongly Disagree (7)}. Descriptive statistics and reliability analysis (using Cronbach’s alpha) were used to assess construct validity and overall user acceptance.

\section{PySelect: A Data-Driven Decision Support System for Software Package Selection}~\label{sec:Framework}

This section presents the development process of the technology selection framework introduced in Figure~\ref{fig:MCDMinSE} and explained in Section~\ref{problemFormulation}, as instantiated for the software package selection problem. The resulting artifact, \textit{PySelect}, operationalizes this framework through a modular architecture designed to support transparent, reproducible, and context-aware technology selection.

The following subsections describe how the conceptual elements of the framework are implemented in practice. The structure of alternatives, features, and quality attributes is realized through automated data pipelines, a knowledge graph that serves as the decision model, and an inference engine that interprets user intent to generate contextually appropriate package recommendations.

\subsection{Data Collection and Extraction Pipelines}\label{sec:data-extraction}

PySelect incorporates three specialized data extraction pipelines that target distinct categories of input: code repositories, package registries, and developer communities. These pipelines collect and transform both structured and unstructured data into standardized representations that populate the Alternatives–Features and Alternatives–Qualities mappings defined in the technology selection framework (see Figure~\ref{fig:MCDMinSE}).

To construct the Alternatives–Features mapping, the system identifies software packages as the alternatives and extracts their relevant characteristics. These features include technical metadata such as \texttt{requires\_python}, keywords, and download statistics from PyPI, as well as developer-defined topics that describe the intended use of each package. Additionally, usage patterns observed in GitHub repositories reveal how these packages are applied in real-world scenarios, offering insight into developer intentions and practical relevance.

The Alternatives–Qualities mapping is derived from developer opinions and community evaluations. To capture this perspective, PySelect collects review content from platforms such as Stack Overflow, Hacker News, G2, and Reddit. Using generative models, these reviews are analyzed to extract sentiment and to map user opinions onto standardized software quality attributes, including maintainability, reliability, and usability.

\begin{figure*}[ht]
    \centering
    \includegraphics[trim=10 0 10 10,width=1.0\textwidth]{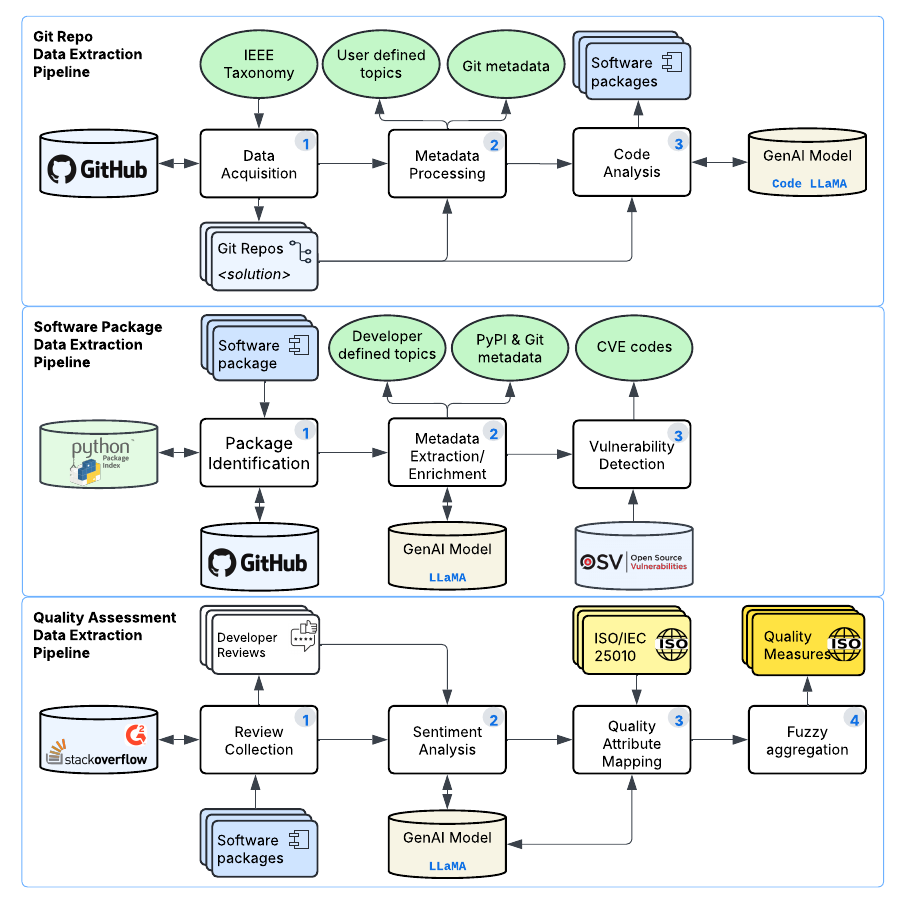}
    \caption{Overview of the data collection and extraction pipelines used in PySelect. The system is composed of three parallel pipelines: (1) the Git Repository Data Extraction Pipeline retrieves relevant repositories from GitHub based on IEEE taxonomy labels and extracts packages and metadata using Code LLaMA; (2) the Software Package Data Extraction Pipeline enriches identified packages with registry-level metadata, developer-defined topics, and vulnerability data from PyPI, GitHub, and the OSV database; and (3) the Quality Assessment Pipeline collects and analyzes developer reviews from community platforms, applies sentiment and quality attribute mapping using Generative AI (LLaMA 3.2 7B), and computes aggregated quality scores through fuzzy logic. Together, these pipelines form the foundation for generating structured, interpretable inputs to PySelect's inference engine.}
    \label{fig:DataExtractionPipelines}
\end{figure*}

Figure~\ref{fig:DataExtractionPipelines} illustrates the architecture of the extraction process, showing the flow of inputs, intermediate transformations, and applied models. The resulting data serves as the foundation for the downstream components of PySelect, including the knowledge graph, inference engine, and recommendation interface. Together, these components enable context-aware and quality-informed analysis of software packages.

\subsubsection{Git Repository Data Extraction Pipeline}

The Git repository pipeline focuses on mining repositories relevant to specific technology domains, using the IEEE taxonomy~\cite{ieee2025taxonomy} as the entry point for topic-driven retrieval. The pipeline proceeds through three sequential stages.

\textbf{(1) Data Acquisition.} This stage begins with a set of IEEE Taxonomy terms as input. In this study, the categories \textit{Computational and Artificial Intelligence} and \textit{Computers and Information Processing} were used to construct GitHub queries. These terms can be extended to include more specific topics based on project needs, such as ``machine learning.'' The query results produce a curated set of GitHub repositories that contain relevant source code. This collection forms the candidate project set, referred to as \texttt{Git Repos <solution>}.

\textbf{(2) Metadata Processing.} The retrieved repositories are analyzed to extract project-level metadata, such as the number of contributors, stars, forks, and update timestamps. In parallel, user-defined topics are inferred through analysis of repository tags, natural language descriptions, and the structural organization of the repository content. These topics support contextual filtering and relevance modeling.

\textbf{(3) Code Analysis.} In the final stage, the source code files within each repository are parsed to identify referenced software packages. A generative AI model (Code LLaMA) is used to extract import statements, dependency references, and context-specific usage patterns. The output is a list of identified packages, each enriched with relevant metadata and contextual information.

\subsubsection{Software Package Data Extraction Pipeline}

This pipeline enriches each identified software package with registry-level metadata obtained from PyPI and GitHub. It adds contextual information derived from developer-defined data and performs a security risk assessment using publicly available vulnerability records.

\textbf{(1) Package Identification.} Using the list of software packages extracted from the Git Repository Data Extraction pipeline, this stage attempts to match each package to its official entry on PyPI. If successful, the corresponding GitHub repository is also identified. Once both sources are linked, the system collects relevant metadata, including version history, maintainer details, release frequency, and repository activity metrics.

\textbf{(2) Metadata Extraction and Enrichment.} The collected metadata is further processed using a generative AI model (LLaMA 3.2 7B) to extract high-level descriptors such as the package’s intended purpose, functional domains, and developer-defined topics. This semantic enrichment supports user intent modeling and enhances the classification and comparison of available alternatives.

\textbf{(3) Vulnerability Detection.} To evaluate potential risks, the pipeline queries the Open Source Vulnerability (OSV) database for known Common Vulnerabilities and Exposures (CVEs) associated with each package. The result is a structured vulnerability profile that informs the quality assessment and helps guide secure package selection.

\subsubsection{Quality Assessment Data Extraction Pipeline}

The third pipeline focuses on capturing developer sentiment and subjective perceptions of software quality. These insights are drawn from community-driven review platforms and are used to assess quality attributes such as reliability, usability, and security.

\textbf{(1) Review Collection.} Developer reviews are retrieved from widely used technical forums and Q\&A platforms, including Stack Overflow, GitHub, Reddit, Hacker News, Dev.to, and CodeProject. Using the package names identified in earlier stages as search queries, the system collects relevant discussion threads, user comments, and informal evaluations.

\textbf{(2) Sentiment Analysis.} Each review is processed using a generative AI model (LLaMA 3.2 7B) to determine sentiment polarity. The model classifies the tone of each statement as positive, negative, or neutral. This step helps filter irrelevant content and prepares the data for quality-oriented aggregation.

\textbf{(3) Quality Attribute Mapping.} The content of each review is then mapped to one or more software quality attributes defined by the ISO/IEC 25010 standard~\cite{iso25010}, such as maintainability, security, and performance. This mapping is also performed by the generative AI model, which interprets the semantic meaning of each statement and aligns it with the appropriate quality attribute.

\textbf{(4) Fuzzy Aggregation.} In the final step, individual sentiment assessments are aggregated into composite quality scores using a fuzzy logic approach~\cite{chen1998aggregating}. Inspired by prior work on architectural knowledge~\cite{ArchitecturePatternsfarshidi2020capturing} and software package evaluation~\cite{hou2024evaluating}, this method classifies sentiment into three categories: Low (L), Medium (M), and High (H). A normalized score between 0 and 1 is computed using the following weighted average:

\[
    \text{Fuzzy Score} = \frac{w_L \cdot L + w_M \cdot M + w_H \cdot H}{L + M + H}
\]

where \(w_L = 0.0\), \(w_M = 0.5\), and \(w_H = 1.0\). The resulting score provides a standardized representation of community sentiment for each quality attribute, which is then used in the ranking and recommendation process.

As an example, consider the case of PyTorch. If developers frequently discuss its performance across forums, the system evaluates the sentiment distribution within those reviews by counting how many are positive, neutral, or negative. This information is then translated into a single quality score that reflects the overall perception of the package's performance. For instance, if the system identifies 100 positive reviews, 10 neutral reviews, and 5 negative reviews related to performance, the resulting score would indicate that PyTorch is perceived to have high performance by the developer community.

\subsection{Decision Model Realization: Knowledge Graph and Inference Engine}\label{sec:knowledge-graph}

To implement the technology selection framework shown in Figure~\ref{fig:MCDMinSE}, PySelect structures the extracted data into a knowledge graph that instantiates the decision model described in Section~\ref{sec:ResearchApproach}. In this model, each software package is represented as an alternative and is linked to associated features and quality attributes. These connections form a semantic network that serves as the foundation for PySelect’s reasoning and recommendation capabilities.

\begin{figure*}[ht]
    \centering
    \includegraphics[trim=0 0 0 0,width=0.9\textwidth]{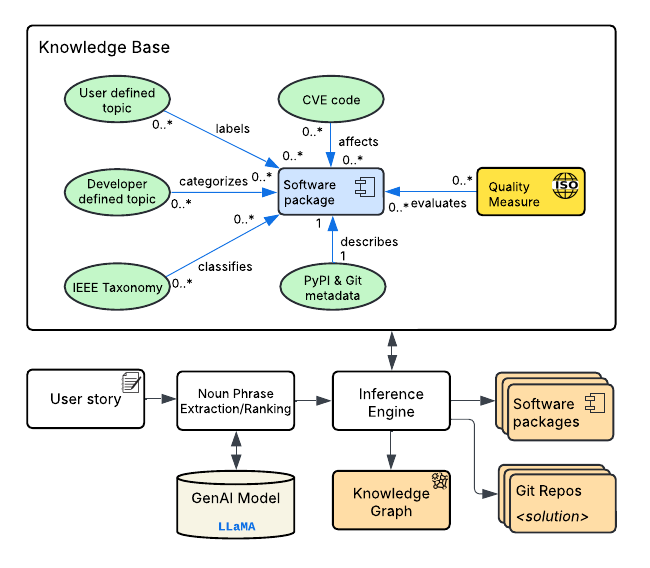}
    \caption{Software Package Inference Engine}
    \label{fig:SoftwarePackageInferenceEngine}
\end{figure*}

Figure~\ref{fig:SoftwarePackageInferenceEngine} presents the architecture of the inference engine and its integration with the knowledge graph. The graph consolidates all outputs from the three data extraction pipelines introduced in Figure~\ref{fig:DataExtractionPipelines}. Each software package is represented as a central node, connected to six types of supporting nodes that capture various evaluation dimensions.

These include user-defined topics and IEEE taxonomy labels extracted from GitHub repositories, developer-defined topics and metadata retrieved from PyPI and GitHub, CVE identifiers indicating known security vulnerabilities, and quality scores derived from community reviews. The user-defined topics are generated by analyzing semantic elements within source code, such as function names, class identifiers, and comments, while developer-defined topics reflect the intended purpose and domain of each package.

Each category of information contributes a distinct perspective to the decision model. User-defined topics represent developer intent, developer-defined topics offer curated descriptions, and IEEE taxonomy labels classify each package within formal domain categories. CVE data informs trustworthiness by linking packages to known vulnerabilities, and quality attributes capture subjective evaluations such as reliability and usability. Descriptive metadata provides technical details, including release history, versioning, and usage metrics.

Most relationships in the knowledge graph are many-to-many. A package may be associated with multiple topics, quality attributes, or vulnerabilities, reflecting the complexity and richness of the software ecosystem. An exception is metadata, which typically links to a single node per versioned release.

\begin{figure*}[ht]
    \centering
    \includegraphics[trim=10 10 10 20,width=1\textwidth]{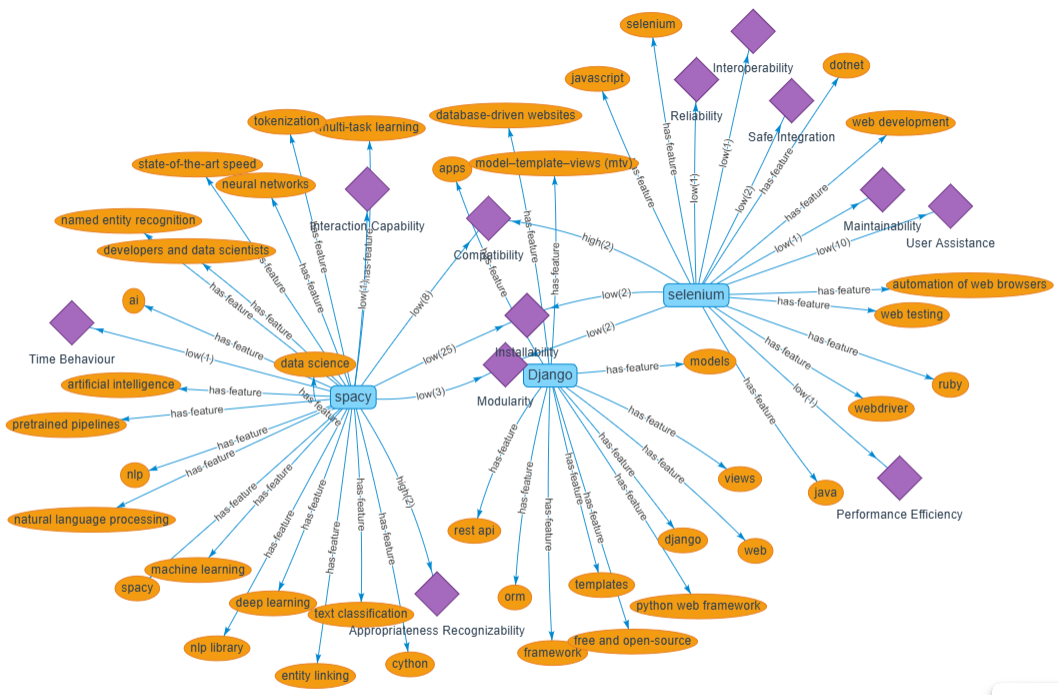}
    \caption{Example subgraph extracted from the knowledge graph showing relationships among three software packages}
    \label{fig:KG}
\end{figure*}

Figure~\ref{fig:KG} illustrates a subgraph extracted from the knowledge graph, focusing on three representative software packages: \texttt{spaCy}, \texttt{Django}, and \texttt{Selenium}. Orange nodes represent supporting data types, such as metadata and topic labels, while purple diamonds correspond to quality dimensions. For clarity, certain node types are grouped and labeled as feature clusters.

The inference engine operates on this graph by processing user-defined requirements. A user begins by submitting a use story that describes the desired functionality, application context, or deployment constraints. This input is parsed by a noun phrase extraction component powered by a generative AI model (LLaMA 3.2 7B), which identifies key terms, extends context, and maps the input to technical vocabulary.

These extracted terms are used to construct a structured query that reflects the user’s intent. The inference engine uses this query to traverse the knowledge graph, identify relevant subgraphs, and retrieve packages that align with the specified constraints. The final output is a ranked list of recommended software packages, each accompanied by contextual metadata and a link to its GitHub repository.

\section{Empirical Evidence}~\label{sec:evaluation}

To evaluate the effectiveness, accuracy, and practical value of PySelect, we conducted a multi-dimensional empirical study incorporating both quantitative and qualitative methods\footnote{All associated research artifacts, including software packages, GitHub repositories, script files, developer reviews, data collection and analysis scripts, experiment results, knowledge base indices, literature study materials, and user study instruments, are available on Mendeley Data~\cite{farshidi2025pypi}. A live version of the PySelect system used during the user study is accessible at \url{https://ai4rse.nl/SoftwarePackageSelection/}.}
. The evaluation spans four key dimensions: (1) technical evaluation of the data collection and extraction pipelines using precision, recall, and coverage metrics; (2) comparative analysis of PySelect’s package recommendations against those generated by leading generative AI tools such as ChatGPT, Copilot, and DeepSeek; (3) a user study grounded in the TAM to assess perceived usefulness, ease of use, attitude, and behavioral intention; and (4) a descriptive analysis of the mined software repositories (MSR) to characterize the structure and content of the PySelect knowledge base, including real-world package usage patterns and ecosystem coverage. This comprehensive approach provides both empirical rigor and developer-centered insights and supports an assessment of PySelect’s utility, reliability, and alignment with real-world development practices.

\subsection{Data Extraction Pipelines}\label{sec:pipeline-eval}
The evaluation of the data extraction pipelines focuses on assessing their accuracy and robustness in collecting relevant information from diverse sources, including GitHub repositories, PyPI metadata, and developer community platforms. Three pipelines were evaluated (see Figure~\ref{fig:DataExtractionPipelines}): (1) Git Repo Data Extraction Pipeline, (2) Software Package Data Extraction Pipeline, and (3) Quality Assessment Data Extraction Pipeline. Each was applied to its corresponding data source, and the outputs were independently reviewed for contextual relevance by two human annotators and two generative AI models (Gemini and GPT). Majority voting~\cite{sheng2008get,penrose1946elementary} was used to determine consensus labels, with human agreement treated as ground truth and Generative AI agreement as the predicted outcome. Precision, recall, F1 score, and accuracy were computed to measure classification performance. The results are reported separately for each pipeline and the specific information extracted.

\subsubsection{Git Repo Data Extraction Pipeline}

The Git Repo Data Extraction Pipeline was evaluated using a statistically representative random sample of 376 GitHub repositories, comprising 845 source code files. This sample size was derived using a 95\% confidence level and a 5\% margin of error, with finite population correction applied to the full dataset of 16,887 repositories.

Two core tasks were assessed: (1) mapping extracted topics to the IEEE taxonomy and (2) identifying user-defined topics from repository content. The second task leveraged an intent code modeling approach, which analyzed descriptive elements such as variable names, function and class identifiers, file names, and inline comments.

\begin{table}[ht]
\centering
\caption{Evaluation of the Git Repo Data Extraction Pipeline}
\begin{tabular}{lcccc}
\toprule
\textbf{Task} & \textbf{Precision} & \textbf{Recall} & \textbf{F1 Score} & \textbf{Accuracy} \\
\midrule
IEEE Taxonomy Mapping      & 0.985 & 0.998 & 0.992 & 0.865 \\
User-defined Topic Extraction & 0.983 & 0.994 & 0.989 & 0.833 \\
\midrule
\# Repositories Evaluated  & \multicolumn{4}{c}{376} \\
\# Source Code Files Analyzed & \multicolumn{4}{c}{845} \\
\bottomrule
\end{tabular}
\label{tab:git-repo-eval}
\end{table}

The pipeline demonstrated high performance across both tasks. For IEEE taxonomy mapping, it achieved a precision of 0.985, recall of 0.998, and F1 score of 0.992. For user-defined topic extraction, it recorded a precision of 0.983, recall of 0.994, and F1 score of 0.989. Accuracy scores were 0.865 and 0.833, respectively, indicating strong alignment with ground truth labels and low misclassification rates. The evaluation results are summarized in Table~\ref{tab:git-repo-eval}.

\subsubsection{Software Package Data Extraction Pipeline}

The Software Package Data Extraction Pipeline was evaluated using a random sample of 374 Python software packages. This sample was calculated based on a population of 12,137 packages, using a 95\% confidence level and a 5\% margin of error, with finite population correction applied. A total of 1,870 textual explanations were processed across the sampled packages.

The evaluation focused on the extraction of developer-defined topics that describe the intended functionality or purpose of the packages. These topics were primarily derived from descriptive fields commonly found in package documentation, including descriptions, summary tags, and other metadata. The extracted topics reflect the developers' own labeling of software functionality.

\begin{table}[ht]
\centering
\caption{Evaluation of the Software Package Data Extraction Pipeline}
\begin{tabular}{lcccc}
\toprule
\textbf{Task} & \textbf{Precision} & \textbf{Recall} & \textbf{F1 Score} & \textbf{Accuracy} \\
\midrule
Developer-defined Topic Extraction & 0.995 & 0.998 & 0.997 & 0.905 \\
\midrule
\# Software Packages Evaluated & \multicolumn{4}{c}{374} \\
\# Textual Explanations Processed & \multicolumn{4}{c}{1,870} \\
\bottomrule
\end{tabular}
\label{tab:package-pipeline-eval}
\end{table}

The pipeline achieved high performance, with a precision of 0.995 and a recall of 0.998, resulting in an F1 score of 0.997. The overall accuracy of 0.905 indicates strong agreement between the predicted labels and the human consensus. These results demonstrate the pipeline’s effectiveness in accurately capturing the topical intent defined by software authors across a diverse range of packages. The evaluation results are presented in Table~\ref{tab:package-pipeline-eval}.

\subsubsection{Quality Assessment Data Extraction Pipeline}

The Quality Assessment Data Extraction Pipeline was evaluated using a sample of 343 Python software packages, drawn from a population of 2,700 records. The sample size was computed using a 95\% confidence level and a 5\% margin of error, applying finite population correction. In total, 3,506 review statements were analyzed from developer-oriented community platforms, such as G2, Stack Overflow, and Hacker News.

Two tasks were evaluated: polarity detection and mapping to ISO/IEC 25010 quality characteristics~\cite{iso25010}. Polarity detection aimed to identify the overall sentiment of each review and achieved strong performance, with a precision of 0.963, recall of 0.959, and F1 score of 0.961. The corresponding accuracy of 0.934 indicates high agreement with human consensus when classifying review tone as positive, negative, or neutral.

In contrast, the task of mapping user statements to ISO/IEC 25010 quality attributes proved more challenging. The pipeline achieved a precision of 0.884 but a substantially lower recall of 0.619, yielding an F1 score of 0.728 and an overall accuracy of 0.669. This discrepancy reflects the difficulty of aligning informal, subjective language with the formal terminology of quality standards. User comments often describe perceived behavior or developer experience in general terms that span multiple dimensions, such as reliability, security, or usability.

\begin{table}[ht]
\centering
\caption{Evaluation of the Quality Assessment Data Extraction Pipeline}
\begin{tabular}{lcccc}
\toprule
\textbf{Task} & \textbf{Precision} & \textbf{Recall} & \textbf{F1 Score} & \textbf{Accuracy} \\
\midrule
Polarity Detection   & 0.963 & 0.959 & 0.961 & 0.934 \\
ISO/IEC 25010 Mapping & 0.884 & 0.619 & 0.728 & 0.669 \\
\midrule
\# Software Packages Evaluated & \multicolumn{4}{c}{343} \\
\# Review Statements Analyzed  & \multicolumn{4}{c}{3,506} \\
\bottomrule
\end{tabular}
\label{tab:quality-pipeline-eval}
\end{table}

Further complicating this task, platforms like Stack Overflow are often focused on technical problem-solving rather than comprehensive software evaluation. Some discussions may be outdated or limited in scope, especially in the context of rapidly evolving software ecosystems. These factors collectively contribute to the reduced performance observed in structured quality attribute mapping. The evaluation results are summarized in Table~\ref{tab:quality-pipeline-eval}.

\subsection{Recommendation Comparison with Generative AI Models}\label{sec:Generative AI-comparison} 

This section evaluates the quality and alignment of PySelect’s recommendations in comparison to those generated by leading generative AI models: ChatGPT, DeepSeek, and GitHub Copilot. The comparison focuses on three metrics: (1) coverage of participant-selected packages, (2) popularity-weighted coverage, and (3) agreement with PySelect’s top-10 ranked recommendations. These metrics collectively assess the effectiveness and contextual alignment of each tool in supporting software package selection.

Table~\ref{tab:pyselect_comparison} presents the results. In terms of raw coverage, DeepSeek achieves the highest score among the Generative AI models, closely followed by ChatGPT and Copilot. When accounting for popularity weighting, all tools perform similarly, with DeepSeek maintaining a slight advantage. Agreement with PySelect’s top-10 recommendations is highest for ChatGPT (0.82), followed by DeepSeek (0.80) and Copilot (0.69).

To assess statistical significance, McNemar’s test~\cite{mcnemar1947note} was applied to binary recommendation outcomes. The results confirm that the recommendation patterns of all Generative AI tools differ significantly from those of PySelect ($p < 0.001$). However, Wilcoxon signed-rank tests~\cite{wilcoxon1992individual} applied to frequency-weighted coverage scores show no statistically significant differences ($p > 0.05$), indicating comparable performance levels when accounting for package usage frequency.

\begin{table}[ht]
\centering
\scriptsize
\caption{Coverage and Agreement of Generative AI Models Compared to PySelect}
\begin{tabular}{p{5.2cm}>{\centering\arraybackslash}p{1.2cm}>{\centering\arraybackslash}p{1.2cm}>{\centering\arraybackslash}p{1.2cm}>{\centering\arraybackslash}p{1.2cm}}
\toprule
\textbf{Evaluation Metric} & \textbf{PySelect} & \textbf{ChatGPT} & \textbf{DeepSeek} & \textbf{Copilot} \\
\midrule
Coverage of participant-selected packages     & 0.91 & 0.88 & 0.97 & 0.91 \\
Coverage (popularity-weighted)                & 0.97 & 0.94 & 0.99 & 0.97 \\
Agreement with PySelect Top-10 recommendations & --   & 0.82 & 0.80 & 0.69 \\
\bottomrule
\end{tabular}
\label{tab:pyselect_comparison}
\end{table}

Table~\ref{fig:PySelectVsGenerative AI} in the appendix illustrates the overlap between user-selected packages and the recommendations produced by PySelect and each Generative AI model. This visual comparison highlights areas of convergence and divergence across different tools and further contextualizes the tabular results.

The PySelect knowledge graph was constructed by analyzing 798{,}669 Python script files from 16{,}887 GitHub repositories, resulting in 39{,}841 distinct software packages. Of these, 18{,}466 (46.4\%) are indexed on PyPI. In contrast, PyPI hosts more than 660{,}000 distinct projects~\cite{pypistats2025}, representing a vast and continuously growing ecosystem. This gap reflects the selective nature of PySelect’s data collection, which focuses on packages actively used in open-source repositories and relevant to IEEE taxonomy categories. As a result, PySelect does not attempt exhaustive coverage of the Python ecosystem but instead prioritizes packages that are both technically meaningful and demonstrably used in real-world development practice.

\subsection{User Study}\label{sec:user-study}

To assess the usability, perceived value, and adoption potential of the PySelect system, we conducted a user study involving software developers and graduate students with relevant programming experience. The study consisted of two parts: a structured questionnaire based on the TAM to quantitatively evaluate perceived usefulness, ease of use, and behavioral intention; and a qualitative feedback session to capture open-ended reflections on system functionality, trustworthiness, and areas for improvement.

\subsubsection{Technology Acceptance Model Evaluation}~\label{sec:tam-study}

To evaluate the acceptance and perceived value of PySelect, a user study was conducted based on the TAM, originally proposed by Davis~\cite{davis1989perceived}. A total of 22 participants engaged in a structured usability session with PySelect, followed by a TAM-based questionnaire comprising 15 items. These items targeted four key constructs: Perceived Usefulness (PU), Perceived Ease of Use (PEOU), Attitude Toward Using (ATT), and Behavioral Intention to Use (BI). Responses were collected using a 7-point Likert scale, where 1 indicated strong agreement and 7 indicated strong disagreement.

Participant selection was designed to reflect a diverse cross-section of the Python developer community. Background data were collected to ensure this diversity, including job roles, educational backgrounds (e.g., computer science degrees, coding bootcamps, self-taught paths), geographic location, years of programming and Python experience, primary use cases (e.g., web development, data analysis, machine learning), preferred development environments, industry domains, employment types, and primary languages. Participants also reported their familiarity with open-source software, use of third-party packages, habits related to developer tools, and comfort with AI-based tools such as ChatGPT and GitHub Copilot.

\begin{table*}[ht]
    \caption{Summary of TAM questionnaire results for PySelect.}
    \centering
    \includegraphics[trim=10 10 10 10,width=0.96\textwidth]{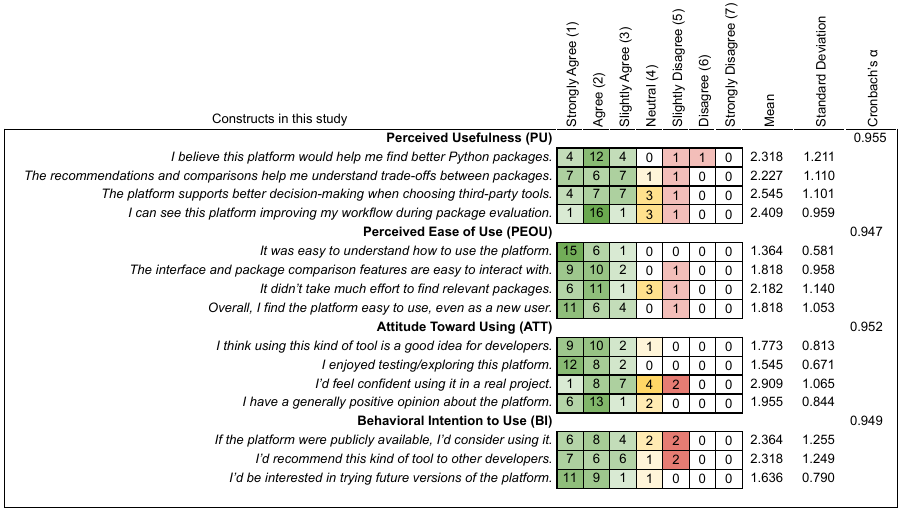}
    \label{tab:tam-results}
\end{table*}

Table~\ref{tab:tam-results} presents the aggregated results, showing item-level response distributions along with corresponding mean scores, standard deviations, and internal reliability coefficients (Cronbach’s $\alpha$~\cite{cronbach1951coefficient}). Across all four constructs, PySelect received consistently favorable evaluations.

\noindent The \textbf{Perceived Usefulness} construct showed strong internal consistency ($\alpha = 0.955$), with mean scores ranging from 2.23 to 2.55. Participants largely agreed that PySelect supports better package decisions, provides helpful trade-off comparisons, and integrates well with their workflow, confirming its value as a decision-support tool.

\noindent \textbf{Perceived Ease of Use} also demonstrated high internal reliability ($\alpha = 0.947$), with all four items receiving low mean scores. For example, the item “ease of understanding” yielded $M = 1.364$, indicating that the platform was perceived as intuitive and easy to navigate. Low standard deviations across items suggest a shared perception of usability.

\noindent The \textbf{Attitude Toward Using} construct ($\alpha = 0.952$) reflected a positive overall perception of the platform. Participants generally viewed PySelect as a good idea and reported enjoying its use. One item, confidence in applying the tool in a real-world project, had a slightly higher mean ($M = 2.909$), suggesting that while the tool is positively received, additional improvements in documentation or robustness may be needed to build full production-level confidence.

\noindent \textbf{Behavioral Intention to Use} scored favorably as well ($\alpha = 0.949$). Participants indicated strong interest in using future versions of PySelect and a willingness to recommend it to others. The item “I’d be interested in trying future versions” received one of the most favorable responses ($M = 1.636$, $SD = 0.790$), indicating sustained engagement potential.

These findings suggest that PySelect is well-received in terms of both perceived utility and usability. The high-reliability coefficients across constructs reinforce the coherence of the questionnaire. While most metrics indicate strong user acceptance, specific areas, such as user confidence, offer direction for future design and communication improvements. These results, although limited by sample size, offer robust initial evidence that PySelect meets the core criteria of a usable and valuable developer support tool.

\subsubsection{Qualitative User Feedback}

A total of 22 participants provided open-ended feedback after using \textit{PySelect}, revealing both commonly appreciated features and recurring suggestions for improvement.

The most frequently cited strength was the \textit{comparison feature}, mentioned by 6 participants (27.3\%). Four participants (18.2\%) emphasized the platform’s \textit{ease of use} and praised its overall \textit{user interface and experience}. The \textit{scoring and ranking system} was highlighted by 3 users (13.6\%) as helpful in identifying relevant packages, while another 3 (13.6\%) credited PySelect with helping them discover \textit{previously unknown or new packages}. Additional positive mentions included \textit{aggregated metadata presentation}, \textit{natural language support}, \textit{semantic search}, and \textit{knowledge graph integration}, each noted by one or two users.

Despite these strengths, participants also provided constructive feedback indicating several areas for enhancement. The most frequent suggestion, raised by 4 users (18.2\%), was to improve the \textit{transparency and interpretability of the scoring system}, particularly through clearer explanations of how scores are assigned. Three participants (13.6\%) requested a broader range of recommendations, while another three expressed concerns about \textit{bias or irrelevance} in the suggested packages. Similarly, 3 users (13.6\%) identified areas for improvement in \textit{UI/UX design}, especially in relation to visual clarity and interactive flow. Other suggestions included the addition of \textit{code usage examples} (2 mentions; 9.1\%), enhanced \textit{filtering and customization options} (2 mentions; 9.1\%), and greater \textit{transparency about data sources} for quality metrics (3 mentions; 13.6\%), such as GitHub, Stack Overflow, or G2. Two participants (9.1\%) also proposed the integration of \textit{LLMs or generative AI} to improve semantic alignment and contextual analysis.

Regarding the \textit{trustworthiness of quality metrics}, 9 participants (40.9\%) expressed full confidence in the reliability of the scoring system, while 4 (18.2\%) reported conditional trust based on transparency and supporting evidence. Three users (13.6\%) were neutral or skeptical, citing the need for better validation and reduced bias. Several participants emphasized that including references or justifications for quality metrics would substantially improve credibility.

In summary, PySelect was positively received for its comparative analysis, ease of use, and scoring features. Nonetheless, user feedback identified several areas for refinement, including the interpretability of scores, breadth of recommendations, interface design, and the transparency of quality assessments.

\subsection{Empirical Analysis of Software Package Usage in Open-Source Repositories} \label{sec:SoftwarePackageUsagePatterns}

This section presents a descriptive analysis of the current state of the PySelect knowledge base at the time of writing. The dataset consists of 798{,}669 Python script files collected from 16{,}887 GitHub repositories, from which 39{,}841 unique software packages were identified. These packages include both widely adopted libraries (e.g., \texttt{numpy}, \texttt{torch}) and lesser-known, domain-specific components. The following analysis characterizes package availability, usage frequency, and domain-specific distribution. These patterns provide empirical context for evaluating the PySelect system and support its underlying design decisions.

\subsubsection{Keyword Frequency Distribution in Domain Features}

To analyze thematic trends within the knowledge base, we examined domain-specific keywords extracted from the metadata. These keywords represent the conceptual and technological focus of the analyzed repositories, covering both general topics such as \textit{machine learning} and \textit{data science}, and more specialized software engineering terms.

A total of 99{,}191 unique keywords were identified, appearing 279{,}563 times. The frequency distribution is highly imbalanced: most keywords appear only once, while a small number recur across many repositories. Table~\ref{tab:union_frequency_distribution} summarizes the distribution across defined frequency intervals.

\begin{table}[htbp]
\centering
\caption{Frequency Distribution of Domain Keywords}
\begin{tabular}{lrr}  
\toprule
\textbf{Frequency Interval} & \textbf{Count} & \textbf{\% of Unique} \\
\midrule
\(x = 1\)                 & 79,997       & 80.65\% \\
\(1 < x \leq 10\)         & 16,546       & 16.68\% \\
\(10 < x \leq 100\)       & 2,353        & 2.37\%  \\
\(100 < x \leq 1{,}000\)  & 283          & 0.29\%  \\
\(1{,}000 < x \leq 10{,}000\) & 12      & 0.01\%  \\
\bottomrule
\end{tabular}
\label{tab:union_frequency_distribution}
\end{table}

Table~\ref{tab:union_top10} lists the most frequently occurring keywords. These reflect the dominant areas of focus within the knowledge base.

\begin{table}[htbp]
\centering
\caption{Top 10 Most Frequent Domain Keywords}
\begin{tabular}{lrr}  
\toprule
\textbf{Keyword} & \textbf{Count} & \textbf{Percentage} \\
\midrule
machine learning           & 4,104 & 1.47\% \\
deep learning              & 1,518 & 0.54\% \\
artificial intelligence    & 1,481 & 0.53\% \\
data science               & 1,273 & 0.46\% \\
computer science           & 1,255 & 0.45\% \\
developer and data scientist & 1,251 & 0.45\% \\
machine intelligence       & 1,232 & 0.44\% \\
data mining                & 1,171 & 0.42\% \\
testing                    & 1,168 & 0.42\% \\
supervised learning        & 1,096 & 0.39\% \\
\bottomrule
\end{tabular}
\label{tab:union_top10}
\end{table}

\noindent
The results highlight the prevalence of artificial intelligence-related topics, including \textit{machine learning}, \textit{deep learning}, and \textit{data mining}. Additionally, keywords such as \textit{testing} and \textit{developer and data scientist} indicate a strong emphasis on practical development concerns.

\subsubsection{Availability of Software Packages on PyPI}

To assess the provenance and distribution of third-party dependencies within the knowledge base, we analyzed the availability of identified packages on PyPI. Of the 39{,}841 unique packages extracted from 798{,}669 script files, 18{,}466 packages (46.35\%) were available on PyPI, while the remaining 21{,}373 packages (53.65\%) were not.

The majority of unavailable packages are likely custom-developed libraries intended for internal use, system-level modules bundled with operating systems, or legacy and forked versions not published to official registries.

This distribution highlights the diversity of real-world software development practices. It also presents a challenge for automated recommendation systems that rely exclusively on PyPI, as such systems may overlook a significant portion of the packages actually used in practice.

\subsubsection{Usage Frequency Distribution of PyPI-Available Packages}

To quantify the adoption of PyPI-available packages within the knowledge base, we analyzed their occurrences across 798{,}669 Python script files. The distribution follows a typical long-tail pattern observed in open-source ecosystems, where the majority of packages are used infrequently and only a few are adopted widely across projects.

As shown in Table~\ref{tab:script_dist_pypi}, 28.72\% of packages are used in only a single script, and 37.98\% appear between two and ten times. An additional 25.25\% are used between eleven and one hundred times. Only 7.18\% occur in more than one hundred scripts. A minimal number, specifically three packages, is referenced in over 100{,}000 files.

\begin{table}[htbp]
    \centering
    \caption{Script File Frequency Distribution for PyPI-Available Packages}
    \begin{tabular}{lrr}
        \toprule
        \textbf{Usage Interval} & \textbf{Package Count} & \textbf{Percentage} \\
        \midrule
        $x = 1$                         & 5,303   & 28.72\% \\
        $1 < x \leq 10$                 & 7,014   & 37.98\% \\
        $10 < x \leq 100$               & 4,662   & 25.25\% \\
        $100 < x \leq 1{,}000$          & 1,326   & 7.18\%  \\
        $1{,}000 < x \leq 10{,}000$     & 133     & 0.72\%  \\
        $10{,}000 < x \leq 50{,}000$    & 21      & 0.11\%  \\
        $50{,}000 < x \leq 100{,}000$   & 4       & 0.02\%  \\
        $x > 100{,}000$                 & 3       & 0.02\%  \\
        \bottomrule
    \end{tabular}
    \label{tab:script_dist_pypi}
\end{table}

This distribution indicates that most packages serve localized or specialized purposes, while a limited set of core libraries is reused extensively. These frequently used packages represent critical dependencies in the Python ecosystem. Issues such as security vulnerabilities, API changes, or deprecation in these libraries can affect a large number of downstream projects. Therefore, recommendation systems should not rely solely on usage popularity but must also consider project-specific relevance and quality indicators when ranking packages.

\subsubsection{Global Package Usage Patterns (All Packages)}

To provide a complete view of package adoption in the knowledge base, we analyzed usage frequency across all 39{,}841 identified packages, including those not available on PyPI. This analysis includes custom-developed, bundled, and externally sourced libraries that are often excluded from registry-based assessments.

\begin{table}[htbp]
    \centering
    \caption{Package Occurrence Frequency Distribution (All Packages)}
    \begin{tabular}{lrr}
        \toprule
        \textbf{Usage Interval} & \textbf{Package Count} & \textbf{Percentage} \\
        \midrule
        $x = 1$                         & 16,601   & 41.67\% \\
        $1 < x \leq 10$                 & 14,505   & 36.41\% \\
        $10 < x \leq 100$               & 6,819    & 17.12\% \\
        $100 < x \leq 1{,}000$          & 1,680    & 4.22\%  \\
        $1{,}000 < x \leq 10{,}000$     & 183      & 0.46\%  \\
        $10{,}000 < x \leq 50{,}000$    & 40       & 0.10\%  \\
        $50{,}000 < x \leq 100{,}000$   & 6        & 0.02\%  \\
        $x > 100{,}000$                 & 5        & 0.01\%  \\
        \bottomrule
    \end{tabular}
    \label{tab:overall_dist}
\end{table}

Over 78\% of all packages are used in fewer than ten scripts, and only 4.8\% appear in more than one hundred. Just eleven packages occur in over 50{,}000 files, indicating that very few libraries achieve widespread adoption.

This pattern reflects the structure of the Python ecosystem, where most packages serve highly specific needs, while a small set of core libraries support a wide range of applications. For software engineering tools focused on dependency analysis, recommendation, or vulnerability management, it is essential to recognize both the critical role of widely used libraries and the variability introduced by specialized components. Effective systems must capture this distribution to remain applicable across diverse development contexts.

\subsubsection{Most Frequently Used PyPI Packages}

To identify the most influential libraries in the knowledge base, we analyzed the ten most frequently imported PyPI-available packages across 798{,}669 Python script files. These packages represent essential tools in domains such as scientific computing, machine learning, testing, and web development.

\begin{table}[htbp]
    \centering
    \caption{Top 10 Most Used PyPI Packages (Relative to Total Scripts)}
    \begin{tabular}{lrr}
        \toprule
        \textbf{Package} & \textbf{Usage Count} & \textbf{Percentage of Scripts (n = 798,669)} \\
        \midrule
        \texttt{numpy}      & 179,815 & 22.51\% \\
        \texttt{typing}     & 133,263 & 16.69\% \\
        \texttt{torch}      & 132,157 & 16.55\% \\
        \texttt{pytest}     & 68,952  & 8.63\%  \\
        \texttt{logging}    & 65,885  & 8.25\%  \\
        \texttt{unittest}   & 59,740  & 7.48\%  \\
        \texttt{time}       & 52,474  & 6.57\%  \\
        \texttt{argparse}   & 39,013  & 4.88\%  \\
        \texttt{datetime}   & 34,920  & 4.37\%  \\
        \texttt{django}     & 34,705  & 4.35\%  \\
        \bottomrule
    \end{tabular}
    \label{tab:top_packages_pypi}
\end{table}

\noindent
The top three packages, \texttt{numpy}, \texttt{typing}, and \texttt{torch}, appear in over 16\% of all scripts, reflecting the central role of numerical computing, type annotations, and deep learning frameworks in Python development. Testing libraries such as \texttt{pytest} and \texttt{unittest}, along with logging utilities, indicate strong community emphasis on software quality. The presence of standard modules like \texttt{argparse}, \texttt{datetime}, and \texttt{time} further shows consistent usage of core language features. The inclusion of \texttt{django} highlights the continued adoption of comprehensive web development frameworks.

The extensive use of these packages makes them critical to a wide range of projects. Changes to their APIs, deprecations, or security vulnerabilities could affect a large portion of the ecosystem. These results underscore the importance of maintaining stability, documentation quality, and backward compatibility in widely adopted libraries.

\section{Discussion}~\label{sec:dicussion}

\subsection{Answering the Research Questions}

The research questions were designed to systematically address the challenges outlined in the problem formulation (Section~\ref{problemFormulation}), specifically the lack of scalable, reproducible, and transparent approaches to third-party software package selection in open-source ecosystems. Each question corresponds to a critical stage in the instantiation and evaluation of the technology selection framework depicted in Figure~\ref{fig:MCDMinSE}. RQ1 focuses on operationalizing the data collection layer by automating the extraction of relevant package metadata, usage patterns, and quality indicators. RQ2 addresses the structuring of this data into a formalized decision model, instantiated as a knowledge graph, and the implementation of an inference mechanism that supports evidence-based reasoning. RQ3 evaluates the effectiveness of the complete framework through quantitative comparison with existing generative AI tools and a user study grounded in the Technology Acceptance Model. Collectively, the research questions serve as a means of validating the framework's applicability, scalability, and practical value for supporting informed software package selection decisions. 

The first question (RQ1) is addressed through the implementation and evaluation of three automated pipelines for data collection and enrichment, described in Section~\ref{sec:data-extraction}. These pipelines extract package usage patterns from 798,669 Python scripts across 16,887 GitHub repositories, normalize metadata from PyPI and GitHub, and assess quality attributes using developer reviews from community platforms. Empirical results demonstrate high extraction accuracy for IEEE taxonomy mapping (F1 = 0.992), user-defined topic identification (F1 = 0.989), and sentiment classification (F1 = 0.961), as reported in Section~\ref{sec:pipeline-eval}. The lower F1 score (0.728) in mapping to ISO/IEC~25010 reflects the inherent difficulty of aligning informal user statements with formal quality models.

The second question (RQ2) is addressed by structuring the collected data into a knowledge graph that encodes the relationships among software packages, usage contexts, quality assessments, vulnerabilities, and semantic topics, as described in Section~\ref{sec:knowledge-graph}. This model supports many-to-many mappings between alternatives, features, and quality attributes, allowing the system to represent complex dependencies and contextual factors. The inference engine interprets user input using noun phrase extraction and traverses the graph to rank packages that align with functional and non-functional requirements. This design integrates MCDM principles with empirical evidence to support context-aware recommendations.

The third question (RQ3) is answered through a multi-dimensional evaluation. A comparative study with three generative AI baselines (ChatGPT, Copilot, DeepSeek) shows that PySelect achieves the highest agreement with participant selections and top-10 recommendation alignment (Section~\ref{sec:Generative AI-comparison}). A user study based on the Technology Acceptance Model, conducted with 22 participants, confirms high levels of Perceived Usefulness (mean = 2.38, $\alpha = 0.955$) and Perceived Ease of Use (mean = 1.81, $\alpha = 0.947$), as detailed in Section~\ref{sec:user-study}. These results indicate that the system is effective in supporting developers during the package selection process.

\subsection{Lessons Learned}

\noindent\textbf{ (1) Software reuse exhibits a long-tail distribution.} The analysis of 798{,}669 Python scripts across 16{,}887 GitHub repositories reveals a highly skewed distribution of software package usage. Among the 39{,}841 unique packages identified, 41.67\% were used only once, and 36.41\% appeared in 2 to 10 scripts. By contrast, only 4.8\% appeared in over 100 scripts, with 11 packages found in more than 50{,}000 files. This usage pattern aligns with previous research on long-tail distributions in software ecosystems. For instance, Bommarito and Bommarito~\cite{bommarito2019empirical} analyzed over 178{,}000 packages and 156 million import statements from PyPI, reporting highly right-skewed distributions in import frequency and contributor counts, explicitly confirming the presence of long-tail patterns in Python package usage. Similarly, Ruohonen et al.~\cite{ruohonen2021large} found that while 46\% of PyPI packages exhibited at least one security issue, these issues were dispersed across a broad set of rarely used packages, suggesting usage patterns dominated by a few popular libraries and a long tail of infrequent ones. These observations are reinforced by Sabbagh~\cite{sabbagh2024vulnerability}, who noted that vulnerabilities and maintenance challenges often cluster in the lower-frequency regions of the dependency graph, where niche or obscure packages reside. Collectively, these studies substantiate the long-tail usage phenomenon observed in this study.

\noindent\textbf{(2) Curated registries are insufficient.} Among all identified packages, 18{,}466 (46.35\%) were indexed on PyPI, while the remaining 53.65\% were not formally published in the central registry. Similar gaps between registry coverage and actual usage have been reported in prior studies. For example, Decan et al.~\cite{decan2019empirical} found that in major ecosystems, only a small number of packages account for most reverse dependencies. This suggests that many packages used in practice are peripheral or not registered. Kula et al.~\cite{kula2017impact} also showed that micro-packages, which are often small or internal modules, make up a large part of dependency graphs. These packages are not always centrally maintained or easy to find in public registries. These findings highlight the limitations of relying only on registry data when selecting or analyzing dependencies.

\noindent\textbf{(3) AI and data-intensive applications dominate the ecosystem.} The analysis of project metadata further reveals that artificial intelligence and data-intensive applications dominate the development landscape. Keywords such as \emph{machine learning} (4{,}104 occurrences), \emph{deep learning} (1{,}518), and \emph{data science} (1{,}273) occur with high frequency. At the same time, terms like \emph{testing} (1{,}168) and \emph{developer and data scientist} (1{,}251) reflect concerns around reliability and software engineering practice. These trends are consistent with prior work. El Aoun et al.~\cite{tidjon2022empirical} report widespread adoption of deep learning libraries such as PyTorch and TensorFlow, often used together within the same projects, reflecting the centrality of AI and data-intensive workflows in contemporary software development. Sculley et al.~\cite{sculley2015hidden} highlight software engineering challenges that arise in machine learning systems, including issues of testing, maintainability, and technical debt, particularly stemming from complex data dependencies and model behavior.

\noindent\textbf{(4) Popularity is not a sufficient factor.} The observed imbalance between widely adopted and niche packages presents unique challenges for dependency recommendation systems. Ranking libraries solely by popularity overlooks contextual suitability, particularly for specialized domains where less common packages may offer more relevant functionality. Kulkarni et al.~\cite{kulkarni2020context} emphasize the value of context-aware recommendation systems that consider domain, functionality, and project-specific needs. Abdalkareem et al.~\cite{abdalkareem2020impact} further report that developers frequently adopt trivial or utility packages not due to popularity but due to perceived reliability. These patterns indicate the necessity of systems capable of discovering and recommending relevant packages beyond those formally indexed in curated registries.

\noindent\textbf{(5) Security must be part of recommendation pipelines.} Security considerations further influence the design of reliable recommendation systems. Some widely adopted packages have been associated with known CVEs, posing long-term risks to dependent projects. Alfadel et al.~\cite{alfadel2023empirical} show that security vulnerabilities in Python packages often persist for extended periods, frequently over three years, and are patched slowly, with many projects delaying updates by several months. These findings underscore the importance of integrating structured vulnerability data from sources such as the OSV database into recommendation models to inform selection decisions with both functionality and risk awareness.

\noindent\textbf{(6) LLMs do not always reflect real-world practice.} An empirical comparison between PySelect and generative AI-based recommendation systems reveals a discrepancy in suggested dependencies. Large language models tend to recommend libraries that are well-documented or broadly visible, but not necessarily those used most frequently in practice. Spracklen et al.~\cite{spracklen2025package} demonstrate that such models can hallucinate non-existent packages, which attackers may exploit by publishing malicious packages under those names. This phenomenon introduces serious supply chain risks. In contrast, PySelect bases its recommendations on large-scale observed usage, thereby aligning more closely with developer behavior and reducing exposure to hallucinated or unsafe dependencies.

\noindent\textbf{(7) Automated mapping from user reviews to quality attributes is still limited.} Interpreting user feedback for structured quality evaluation remains a complex task. While sentiment classification models achieve strong performance, as evidenced in this study with an F1 score of 0.961, the mapping of reviews to formal ISO/IEC~25010 quality attributes achieved a lower score of 0.728. This reflects broader challenges observed in the literature. Jabborov et al.~\cite{jabborov2023taxonomy} found that most quality assessment frameworks for intelligent systems lack automation and require manual intervention when mapping qualitative indicators, such as user sentiment, to structured standards. As a result, effective use of user-generated content for quality assessment may depend on combining natural language processing with semi-automated or curated approaches.

\noindent\textbf{(8) Developer trust is multifaceted.} Software package selection decisions are influenced by a combination of evidence-based factors rather than documentation quality or project visibility alone. Developers consider real-world usage frequency as an indicator of stability and community endorsement~\cite{bogart2015breaks, abdalkareem2020impact}. Known security vulnerabilities, especially those disclosed in public databases, can significantly reduce trust, even for popular libraries~\cite{alfadel2023empirical}. Community sentiment and user reviews also shape trust perceptions, providing social cues about usability, reliability, and active maintenance~\cite{alfadel2023empirical, jabborov2023taxonomy}. Additionally, contextual relevance, how well a package fits the specific functionality, domain, or project constraints, plays a critical role in selection decisions~\cite{kulkarni2020context}. These diverse factors interact in nuanced ways, suggesting that trust in dependencies emerges from a blend of technical evidence, social validation, and contextual appropriateness rather than from any single dominant factor.

\subsection{Threats to Validity}~\label{sec:ThreatstoValidity}

This section discusses potential limitations of the study across four standard dimensions of validity: construct, internal, external, and reliability~\cite{wohlin2012experimentation}.

\textbf{Construct Validity.} This study combined technical evaluation and user perception to assess the effectiveness of the proposed system. To evaluate perceived usefulness and usability, we used the TAM~\cite{davis1989perceived} with a standardized 15-item instrument. Internal consistency across TAM constructs was high, with Cronbach’s alpha values above 0.94~\cite{cronbach1951coefficient}. Participants were known to the researchers and took part in guided interviews. To preserve integrity, each participant was informed that responses would remain anonymous and that no identifying information would be reported. Although personal familiarity may introduce bias, the structured questionnaire and consistent interview protocol helped minimize interpretation variance. In the technical pipeline, the F1 score of 0.728 for mapping reviews to ISO/IEC~25010 attributes highlights limitations in interpreting informal developer language. This risk was addressed using LLM-based topic mapping, filtered sentiment analysis, and fuzzy aggregation to increase interpretability.

\textbf{Internal Validity.} The data extraction pipelines were validated using statistically representative samples based on finite population correction. Ground truth labels were generated through majority voting involving two human annotators and two LLMs. While this improves robustness, misclassification in topic or sentiment extraction may still affect downstream accuracy. To reduce this risk, disagreement cases were flagged and manually reviewed. In the user study, participants’ prior familiarity with specific packages may have influenced preference judgments. To incorporate realistic contexts, users were asked to provide examples of script files they had previously developed and to explain their rationale for selecting specific packages. These discussions allowed clarification of task objectives while avoiding leading prompts.

\textbf{External Validity.} The knowledge base was constructed from GitHub-hosted Python repositories, which may not capture practices in closed-source, industrial, or regulated environments. Of the 39{,}841 packages identified, 53.65\% were not indexed on PyPI, reflecting reliance on system-level or custom modules. Although this limits generalizability to curated ecosystems, the inclusion of non-indexed packages and real usage data improves the representativeness of the dataset. The user study involved 22 participants from academic and developer communities with varying backgrounds. While the sample was diverse in expertise, its size and recruitment scope limit broad generalization. Stratified participant roles and domain profiling were used to mitigate this issue.

\textbf{Reliability.} The system architecture, data pipelines, and evaluation procedures were implemented as reproducible scripts with fixed seeds and deterministic behavior. External data sources such as GitHub, PyPI, OSV, and Stack Overflow introduce potential variability due to changes in APIs, rate limits, or schema evolution. To address this, data was cached during collection, failures were logged, and validation checks were applied across pipeline stages. While the system can be replicated using the published artifacts, slight discrepancies may arise if upstream data or tool behavior changes.

\section{Related Work} \label{sec:RelatedWork}

A wide range of decision support approaches has been proposed for the selection of Commercial-Off-The-Shelf (COTS) components, such as third-party software packages. These studies vary in their methodological strategies, including publication time span, application domain, decision-making logic, data acquisition and modeling approaches, and evaluation techniques. To systematically analyze this landscape, we conducted a structured snowballing review. A total of 52 peer-reviewed studies were identified and categorized along 13 methodological dimensions. The result is summarized in Table~\ref{fig:LS}, which presents a comparative matrix of the reviewed literature.

\begin{table}[!ht]
    \caption{Overview of related work identified through a snowballing literature review on COTS (e.g., software packages) selection. The studies are categorized by: publication year; decision-making domain; user intent modeling techniques; number of decision alternatives and criteria; applied decision-making methods; data collection sources; data collection strategies (manual, semi-automated, automated); employed software quality models; weighting techniques; and ranking methods.}
    \centering
    \includegraphics[width=1.\textwidth]{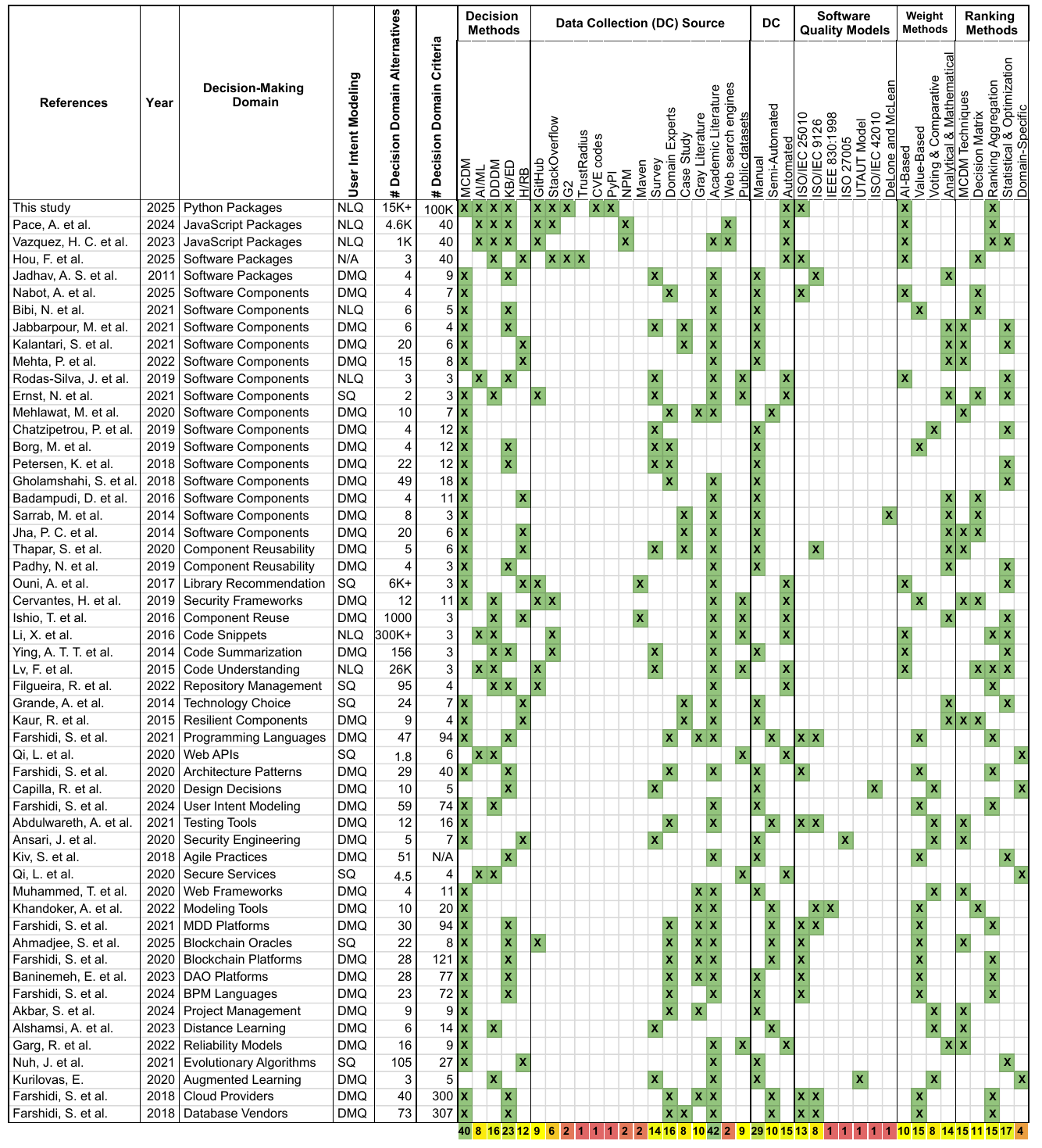}
    \label{fig:LS}
\end{table}

\subsection{Decision-Making Domains and Scope}
The reviewed studies, covering the period from 2011 to 2025, span various decision-making domains, including software package selection~\cite{hou2024evaluating,jadhav2011framework}, software component selection~\cite{nabot2025utilizing,bibi2021conceptual,jabbarpour2021framework}, and JavaScript package evaluation~\cite{pace2023javascript,vazquez2023recommender}, which are closely related to the domain addressed in this study. Some of the studies (as a subset of those summarized in Table~\ref{fig:LS}) specifically focus on the selection of reusable components from repositories such as Maven~\cite{ouni2017search,ishio2016software} and npm~\cite{pace2023javascript,vazquez2023recommender}. In contrast, no peer-reviewed high-quality publications were identified in the context of the PyPI ecosystem. Additional decision-making domains found in the literature (also as subsets of Table~\ref{fig:LS}) include cloud service selection~\cite{Cloudfarshidi2018decision,Serverlesshamza2025decision}, blockchain platforms~\cite{Blockchainfarshidi2020decision}, business process modeling tools~\cite{BPMLfarshidi2024business}, testing tools~\cite{abdulwareth2021toward}, and project management~\cite{akbar2023multi}. The "Decision-Making Domain" column in Table~\ref{fig:LS} provides an overview of the thematic areas identified across these studies.

\subsection{User Intent Modeling Approaches}
The reviewed studies employed various methods for eliciting user requirements and modeling user intent, as summarized in the \textit{User Intent Modeling} column of Table~\ref{fig:LS}. One common approach is natural language queries (NLQ)~\cite{pace2023javascript, vazquez2023recommender, li2016relationship}, where decision-makers define their preferences using natural language inputs. PySelect follows this approach by allowing users to submit user stories written in natural language, which are then parsed and transformed into structured decision queries (see Figure~\ref{fig:SoftwarePackageInferenceEngine}). Another subset of studies utilized decision-making queries (DMQ), including pairwise comparisons and prioritization schemes~\cite{jadhav2011framework, nabot2025utilizing, khandoker2022towards, farshidi2021model}. These methods are typically integrated into systems that apply multi-criteria decision-making techniques and rely on explicit user preferences structured around predefined attributes. Additionally, structured query (SQ) interfaces, similar to SQL, have been applied to enable users to retrieve alternatives based on formal criteria and constraints~\cite{qi2020spatial, Oracleahmadjee2025decision, qi2020data, grande2014framework}. While these methods offer precision, they often assume technical proficiency and may not be accessible to all users.PySelect distinguishes itself by integrating NLQ-based intent modeling with MCDM logic, enabling accessibility for non-expert users while maintaining formal decision support grounded in empirical data.

\subsection{Decision Alternatives and Criteria}
The number of decision alternatives considered across the reviewed studies varies substantially. Some approaches operate on relatively small sets of alternatives (e.g., 2–30 options), while others address large-scale scenarios involving thousands or even hundreds of thousands of candidates. Notable examples include studies evaluating over 6{,}000~\cite{ouni2017search}, 26{,}000~\cite{lv2015codehow}, and 300{,}000~\cite{li2016relationship} alternatives. Evaluation complexity also differs considerably: most studies define between 3 and 20 decision criteria, though a few employ highly detailed models with more than 90 criteria~\cite{farshidi2021decision,farshidi2021model}. Table~\ref{fig:LS}, columns six and seven, summarizes the number of alternatives and criteria used in each study. In domains such as software package selection, particularly in large open-source ecosystems, the number of alternatives can scale exponentially. PySelect addresses this scalability challenge by supporting over 100{,}000 alternatives, significantly exceeding the range observed in most prior work. While the number of alternatives is large, the number of decision criteria is often limited due to manual specification; in contrast, PySelect integrates automated pipelines to derive a broader set of criteria from empirical data sources.

\subsection{Decision-Making Methods and Logic}
The reviewed studies demonstrate a spectrum of decision-making complexity, ranging from lightweight heuristics to comprehensive multi-attribute evaluation frameworks. Column eight of Table~\ref{fig:LS} summarizes the \textit{Decision-Making Methods} employed. Multi-Criteria Decision-Making (MCDM) techniques are the most prevalent, appearing in 39 studies. These models either explicitly adopt MCDM formalisms or implicitly follow multi-attribute evaluation principles. Within the MCDM family, classical approaches such as Analytic Hierarchy Process (AHP), Technique for Order Preference by Similarity to Ideal Solution (TOPSIS), and other structured evaluation techniques are widely used. These are categorized as heuristic and rule-based methods (H/RB), found in 12 studies~\cite{kalantari2021optimal, mehta2022integration, badampudi2016software, thapar2020quantifying, padhy2019identifying, nuh2021performance}. Knowledge-based and expert-driven (KB/ED) approaches are employed in 22 studies, where domain knowledge or expert input informs the decision process~\cite{cervantes2019data, ishio2016software, li2016relationship, ying2014selection, lv2015codehow}. Data-driven decision-making (DDDM) methods appear in 15 studies, typically leveraging large-scale empirical evidence extracted from software repositories or usage patterns~\cite{qi2020data, filgueira2022inspect4py, vazquez2023recommender, qi2020spatial}. Additionally, artificial intelligence and machine learning-based (AI/ML) systems are utilized in 7 studies~\cite{qi2020data, qi2020spatial, rodas2019resdec, li2016relationship, lv2015codehow}, where learning algorithms contribute to automated recommendation or classification. PySelect employs a hybrid decision-making strategy that combines elements of MCDM, AI/ML, DDDM, and KB/ED. This integration enables both scalable automation and knowledge-informed reasoning, allowing the system to model structured evaluation logic while adapting to empirical evidence and user intent.

\subsection{Data Collection Sources and Strategies}
The reviewed studies utilize a wide range of data collection (DC) sources, as summarized in the corresponding column of Table~\ref{fig:LS}. Academic literature is the most commonly used source, referenced in 42 studies~\cite{kaurfuzzy, farshidi2021decision, ArchitecturePatternsfarshidi2020capturing, IntentModelfarshidi2024understanding, abdulwareth2021toward}. Domain expert input, typically gathered through structured or semi-structured interviews, appears in 16 studies~\cite{gholamshahi2019software, farshidi2021decision, Cloudfarshidi2018decision, Databasefarshidi2018decision}. Surveys were used in 14 cases to gather stakeholder feedback or practitioner preferences~\cite{rodas2019resdec, ernst2019component, chatzipetrou2020component, borg2019selecting, petersen2017choosing}, while gray literature, including technical documentation, vendor specifications, and online articles, served as a data source in 10 studies~\cite{farshidi2021decision, muhammed2020selecweb, khandoker2022towards, farshidi2021model, Oracleahmadjee2025decision}. In addition to traditional sources, some studies leveraged online repositories and community platforms. GitHub was used in 8 studies~\cite{ouni2017search, cervantes2019data, lv2015codehow, filgueira2022inspect4py}, and Stack Overflow in 5~\cite{pace2023javascript, hou2024evaluating, cervantes2019data, li2016relationship, ying2014selection}. Other sources include software review platforms such as G2 and TrustRadius~\cite{hou2024evaluating}, as well as publicly available datasets~\cite{qi2020data, qi2020spatial, rodas2019resdec, ernst2019component}. While some studies integrate multiple heterogeneous data sources, none were found to incorporate structured vulnerability data, such as CVE codes, into the decision-making process for software package selection. In contrast, PySelect introduces this additional source by leveraging CVE records to assess security risks, thereby extending the scope and granularity of evidence-based evaluation.

The reviewed studies exhibit varying levels of automation in their data collection strategies, as outlined in the corresponding column of Table~\ref{fig:LS}. Manual data collection dominates the literature, employed in 29 studies~\cite{kaurfuzzy, ArchitecturePatternsfarshidi2020capturing, capilla2020teaching, IntentModelfarshidi2024understanding}, where data is typically gathered through expert input, document analysis, or manual coding. Semi-automated approaches, involving techniques such as web crawling or metadata extraction followed by human validation, are found in 10 studies~\cite{farshidi2021decision, abdulwareth2021toward, khandoker2022towards}. Fully automated data collection pipelines are employed in 14 studies~\cite{pace2023javascript, vazquez2023recommender, hou2024evaluating, rodas2019resdec, ernst2019component}, often relying on programmatic access to repositories, community platforms, or public APIs. PySelect belongs to the latter category, implementing fully automated data pipelines that integrate information from software repositories, package registries, developer forums, and vulnerability databases, as detailed in Section~\ref{sec:data-extraction}. This level of automation enables scalability and reproducibility in the software package selection process, distinguishing PySelect from most prior work.

\subsection{Quality Models, Weighting, and Ranking Techniques}
A wide range of software quality models is applied across the reviewed studies, as summarized in the relevant column of Table~\ref{fig:LS}. ISO/IEC 25010~\cite{iso25010} is the most frequently referenced standard, appearing in 12 studies~\cite{hou2024evaluating, nabot2025utilizing, thapar2020quantifying, farshidi2021decision, ArchitecturePatternsfarshidi2020capturing, abdulwareth2021toward}. ISO/IEC 9126, an earlier quality model, is used in 8 studies~\cite{jadhav2011framework, farshidi2021decision, abdulwareth2021toward, Cloudfarshidi2018decision, Databasefarshidi2018decision}. Additional standards include IEEE 830:1998~\cite{khandoker2022towards}, ISO/IEC 42010~\cite{capilla2020teaching}, ISO 27005~\cite{ansari2020fuzzy}, the Unified Theory of Acceptance and Use of Technology (UTAUT)~\cite{kurilovas2020data}, and the DeLone and McLean information systems success model~\cite{sarrab2014empirical}. PySelect applies ISO/IEC 25010 as its foundational quality model, leveraging it to interpret and structure quality-related factors extracted from developer-generated content. As detailed in Section~\ref{sec:data-extraction}, the system maps community reviews to ISO/IEC 25010 attributes using generative models, thereby incorporating user-perceived software quality into its recommendation process.

Weighting techniques employed across the reviewed studies vary in terms of formalism and complexity, as shown in the \textit{Weighting Methods} column of Table~\ref{fig:LS}. AI-based weight derivation is applied in 9 studies~\cite{hou2024evaluating, pace2023javascript, li2016relationship, ying2014selection, vazquez2023recommender}, typically relying on machine learning or natural language models to infer criteria importance from user input or historical data. Value-based strategies, which prioritize criteria based on domain relevance or expert judgment, are used in 15 studies~\cite{farshidi2021decision, ArchitecturePatternsfarshidi2020capturing, IntentModelfarshidi2024understanding, kiv2018agile, khandoker2022towards, Blockchainfarshidi2020decision}. Comparative and voting-based approaches, often grounded in consensus-building or pairwise comparisons, are found in 8 studies~\cite{capilla2020teaching, akbar2023multi, alshamsi2023multi, kurilovas2020data}. Analytical and mathematically grounded techniques, such as those based on utility theory or normalization models, appear in 14 studies~\cite{jadhav2011framework, jabbarpour2021framework, kalantari2021optimal, mehta2022integration, kaurfuzzy, garg2021decision}. PySelect applies an AI-based weighting strategy using a generative language model (LLaMA), which interprets user stories and extracts feature priorities based on semantic relevance (see Section~\ref{fig:SoftwarePackageInferenceEngine}). This approach enables adaptive, context-aware weighting without requiring manual input from the decision maker.

A diverse set of ranking techniques is applied across the reviewed studies, as presented in the \textit{Ranking Methods} column of Table~\ref{fig:LS}. MCDM-specific techniques such as AHP and TOPSIS are employed in 15 studies~\cite{jabbarpour2021framework, kalantari2021optimal, mehta2022integration, mehlawat2020multi, jha2014optimal, thapar2020quantifying}. Decision matrices, which score and compare alternatives based on weighted criteria, appear in 11 studies~\cite{hou2024evaluating, bibi2021conceptual, ernst2019component, khandoker2022towards, kaurfuzzy}. Ranking aggregation methods, used to consolidate multiple criteria or scoring dimensions into a unified ranking, are applied in 14 studies~\cite{Blockchainfarshidi2020decision, BPMLfarshidi2024business, pace2023javascript, filgueira2022inspect4py, vazquez2023recommender}. Statistical and optimization-based approaches, such as regression models or utility maximization, appear in 17 studies~\cite{kalantari2021optimal, li2016relationship, nuh2021performance, chatzipetrou2020component}. Domain-specific techniques, tailored to particular contexts or evaluation goals, are used less frequently but are present in selected studies~\cite{kurilovas2020data, qi2020spatial, qi2020data, capilla2020teaching}. PySelect applies a ranking aggregation strategy to synthesize quality indicators, usage patterns, and contextual relevance into a unified score. This enables transparent presentation of ranked alternatives to decision-makers based on multidimensional evidence.

\section{Conclusion and Future Work}~\label{sec:conclusion}

In this study, we proposed a technology selection framework for software engineering based on multi-criteria decision-making and empirical data integration. To operationalize this framework, we developed \textit{PySelect}, a data-driven decision support system for selecting Python software packages. PySelect combines automated data pipelines, a structured knowledge graph, and an inference engine to generate context-aware recommendations aligned with project-specific requirements.

The system was applied to a large-scale dataset of 798,669 Python scripts from 16,887 GitHub repositories, resulting in the identification of 39,841 unique software packages. The analysis revealed a long-tail distribution of package usage, highlighting the need for structured selection mechanisms beyond popularity-based approaches. Evaluation of the data pipelines showed high accuracy in metadata extraction, topic modeling, and sentiment-based quality assessment. A user study based on the Technology Acceptance Model indicated strong acceptance in terms of perceived usefulness and ease of use.

Looking ahead, several avenues of future work are envisioned. First, we plan to extend the analysis to other ecosystems (e.g., JavaScript/NPM or R/CRAN) to evaluate the generalizability of our approach. Second, enhancements to the system could include distinguishing between runtime and development dependencies, integrating vulnerability scanning tools, and providing dynamic quality indicators such as documentation completeness or test coverage. Third, while our user study focused on perceived usefulness, future evaluations could include longitudinal studies or controlled experiments to measure the impact of PySelect on task efficiency, decision accuracy, and software quality.

\section*{Acknowledgements}

This work was funded through seed funding in the context of the AgriDash project, a collaboration among Wageningen University \& Research, Eindhoven University of Technology, and Utrecht University in the Netherlands. It was conducted as part of the AI4RSE Lab’s initiatives\footnote{\url{https://ai4rse.nl/}} within the Information Technology Group at Wageningen University \& Research. The authors would like to thank all members of the AI4RSE Lab for their valuable feedback and discussions that helped shape the development of the \textit{PySelect} system. We also appreciate the efforts of the anonymous reviewers, whose thoughtful comments and suggestions have helped improve the quality and clarity of this work.

\appendix
\section{User Study}

This appendix provides additional material related to the user study conducted to evaluate the usefulness and relevance of \textit{PySelect}. As part of the study, participants were asked to name Python packages they had used recently and to describe the purpose of the package, the reason for their selection (such as popularity, documentation quality, or ease of use), and any alternative packages they were aware of.

These purpose–rationale combinations were then used to formulate structured queries. Each query was submitted to \textit{PySelect} as well as to three generative AI models: ChatGPT, GitHub Copilot, and DeepSeek. The aim was to assess whether the tools could return recommendations that aligned with the user’s stated goals and reasoning.

Table~\ref{fig:PySelectVsGenerative AI} visualizes the comparison. For each use case, such as dataframe management or machine learning, the figure shows which tools returned relevant packages and whether those recommendations reflected the user’s rationale. The comparison highlights differences in how each system interprets intent and supports software package selection.

These results are consistent with the findings in Section~\ref{sec:user-study}, showing that \textit{PySelect} provides recommendations that are more often aligned with stated rationale and purpose than those offered by general-purpose generative models.

\begin{table*}[!ht]
    \caption{Comparison of model responses based on user-reported package selection rationales. The figure shows whether \textit{PySelect} and generative AI models returned relevant package suggestions based on user-stated purposes and reasons such as documentation quality or popularity.}
    \centering
    \includegraphics[trim=50 10 50 10,width=0.63\textwidth]{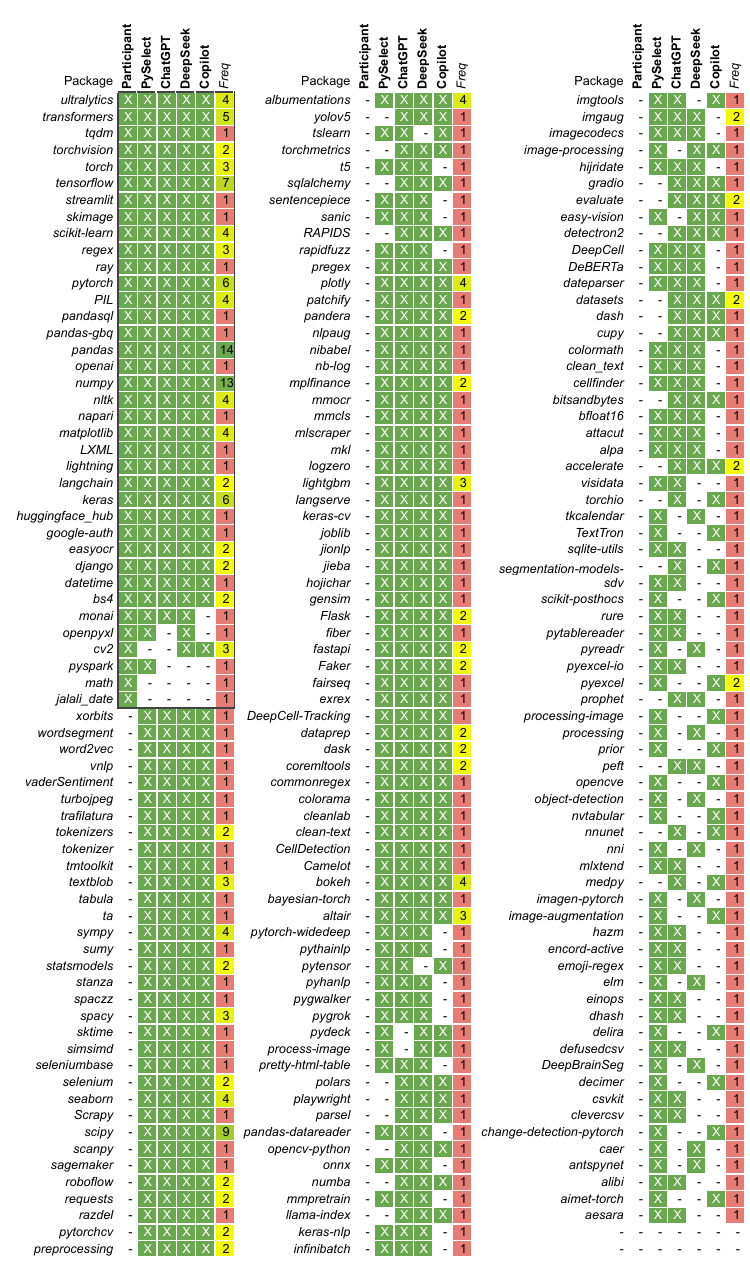}
    \label{fig:PySelectVsGenerative AI}
\end{table*}

\newpage
\bibliographystyle{spbasic}   
\bibliography{references}     

\begin{thebibliography}{110}
\providecommand{\natexlab}[1]{#1}
\providecommand{\url}[1]{{#1}}
\providecommand{\urlprefix}{URL }
\expandafter\ifx\csname urlstyle\endcsname\relax
  \providecommand{\doi}[1]{DOI~\discretionary{}{}{}#1}\else
  \providecommand{\doi}{DOI~\discretionary{}{}{}\begingroup \urlstyle{rm}\Url}\fi
\providecommand{\eprint}[2][]{\url{#2}}

\bibitem[{Abdalkareem et~al.(2020)Abdalkareem, Oda, Mujahid, and Shihab}]{abdalkareem2020impact}
Abdalkareem R, Oda V, Mujahid S, Shihab E (2020) On the impact of using trivial packages: An empirical case study on npm and pypi. Empirical Software Engineering 25(2):1168--1204

\bibitem[{Abdulwareth and Al-Shargabi(2021)}]{abdulwareth2021toward}
Abdulwareth AJ, Al-Shargabi AA (2021) Toward a multi-criteria framework for selecting software testing tools. IEEE Access 9:158872--158891

\bibitem[{Afroogh et~al.(2024)Afroogh, Akbari, Malone, Kargar, and Alambeigi}]{afroogh2024trust}
Afroogh S, Akbari A, Malone E, Kargar M, Alambeigi H (2024) Trust in ai: progress, challenges, and future directions. Humanities and Social Sciences Communications 11(1):1--30

\bibitem[{Ahmadjee et~al.(2025)Ahmadjee, Mera-G{\'o}mez, Farshidi, Bahsoon, and Kazman}]{Oracleahmadjee2025decision}
Ahmadjee S, Mera-G{\'o}mez C, Farshidi S, Bahsoon R, Kazman R (2025) Decision support model for selecting the optimal blockchain oracle platform: An evaluation of key factors. ACM Transactions on Software Engineering and Methodology 34(1):35

\bibitem[{Akbar et~al.(2023)Akbar, Ullah, Khan, Asghar, Zubair, and Zheng}]{akbar2023multi}
Akbar S, Ullah R, Khan R, Asghar I, Zubair M, Zheng Z (2023) A multi-criteria decision-making framework for software project management tool selection. In: Proceedings of the 2023 9th International Conference on Computer Technology Applications, pp 184--191

\bibitem[{Akhavani et~al.(2025)Akhavani, Ousat, and Kharraz}]{akhavani2025open}
Akhavani SA, Ousat B, Kharraz A (2025) Open source, open threats? investigating security challenges in open-source software. arXiv preprint arXiv:250612995

\bibitem[{Alfadel et~al.(2023)Alfadel, Costa, and Shihab}]{alfadel2023empirical}
Alfadel M, Costa DE, Shihab E (2023) Empirical analysis of security vulnerabilities in python packages. Empirical Software Engineering 28(3):59

\bibitem[{Alshamsi et~al.(2023)Alshamsi, El-Kassabi, Serhani, and Bouhaddioui}]{alshamsi2023multi}
Alshamsi AM, El-Kassabi H, Serhani MA, Bouhaddioui C (2023) A multi-criteria decision-making (mcdm) approach for data-driven distance learning recommendations. Education and Information Technologies 28(8):10421--10458

\bibitem[{Ansari et~al.(2020)Ansari, Al-Zahrani, Pandey, and Agrawal}]{ansari2020fuzzy}
Ansari MTJ, Al-Zahrani FA, Pandey D, Agrawal A (2020) A fuzzy topsis based analysis toward selection of effective security requirements engineering approach for trustworthy healthcare software development. BMC Medical Informatics and Decision Making 20:1--13

\bibitem[{Badampudi et~al.(2016)Badampudi, Wohlin, and Petersen}]{badampudi2016software}
Badampudi D, Wohlin C, Petersen K (2016) Software component decision-making: In-house, oss, cots or outsourcing-a systematic literature review. Journal of Systems and Software 121:105--124

\bibitem[{Baninemeh et~al.(2023)Baninemeh, Farshidi, and Jansen}]{DAObaninemeh2023decision}
Baninemeh E, Farshidi S, Jansen S (2023) A decision model for decentralized autonomous organization platform selection: Three industry case studies. Blockchain: Research and Applications 4(2):100127

\bibitem[{Bavota et~al.(2015)Bavota, Canfora, Di~Penta, Oliveto, and Panichella}]{bavota2015apache}
Bavota G, Canfora G, Di~Penta M, Oliveto R, Panichella S (2015) How the apache community upgrades dependencies: an evolutionary study. Empirical Software Engineering 20(5):1275--1317

\bibitem[{Berntsson et~al.(2017)Berntsson, Strand{\'e}n, and Warg}]{berntsson2017evaluation}
Berntsson PS, Strand{\'e}n L, Warg F (2017) Evaluation of open source operating systems for safety-critical applications. In: International Workshop on Software Engineering for Resilient Systems, Springer, pp 117--132

\bibitem[{Bibi et~al.(2021)Bibi, Rana, Naseer et~al.}]{bibi2021conceptual}
Bibi N, Rana T, Naseer A, et~al. (2021) Conceptual model for component selection: A research review on existing techniques. In: 2021 International Conference on Communication Technologies (ComTech), IEEE, pp 22--27

\bibitem[{Bogart et~al.(2015)Bogart, K{\"a}stner, and Herbsleb}]{bogart2015breaks}
Bogart C, K{\"a}stner C, Herbsleb J (2015) When it breaks, it breaks: How ecosystem developers reason about the stability of dependencies. In: 2015 30th IEEE/ACM International Conference on Automated Software Engineering Workshop (ASEW), IEEE, pp 86--89

\bibitem[{Bommarito and Bommarito(2019)}]{bommarito2019empirical}
Bommarito E, Bommarito M (2019) An empirical analysis of the python package index (pypi). arXiv preprint arXiv:190711073

\bibitem[{Borg et~al.(2019)Borg, Chatzipetrou, Wnuk, Al{\'e}groth, Gorschek, Papatheocharous, Shah, and Axelsson}]{borg2019selecting}
Borg M, Chatzipetrou P, Wnuk K, Al{\'e}groth E, Gorschek T, Papatheocharous E, Shah SMA, Axelsson J (2019) Selecting component sourcing options: a survey of software engineering’s broader make-or-buy decisions. Information and Software Technology 112:18--34

\bibitem[{Capilla et~al.(2020)Capilla, Zimmermann, Carrillo, and Astudillo}]{capilla2020teaching}
Capilla R, Zimmermann O, Carrillo C, Astudillo H (2020) Teaching students software architecture decision making. In: Software Architecture: 14th European Conference, ECSA 2020, L'Aquila, Italy, September 14--18, 2020, Proceedings 14, Springer, pp 231--246

\bibitem[{Cervantes et~al.(2019)Cervantes, Kazman, Ryoo, Cho, Cho, Kim, and Kang}]{cervantes2019data}
Cervantes H, Kazman R, Ryoo J, Cho J, Cho G, Kim H, Kang J (2019) Data-driven selection of security application frameworks during architectural design. In: Proceedings of the 52nd Hawaii International Conference on System Sciences

\bibitem[{Chatzipetrou et~al.(2020)Chatzipetrou, Papatheocharous, Wnuk, Borg, Al{\'e}groth, and Gorschek}]{chatzipetrou2020component}
Chatzipetrou P, Papatheocharous E, Wnuk K, Borg M, Al{\'e}groth E, Gorschek T (2020) Component attributes and their importance in decisions and component selection. Software quality journal 28:567--593

\bibitem[{Chen(1998)}]{chen1998aggregating}
Chen SM (1998) Aggregating fuzzy opinions in the group decision-making environment. Cybernetics \& Systems 29(4):363--376

\bibitem[{Choudhuri et~al.(2025)Choudhuri, Trinkenreich, Pandita, Kalliamvakou, Steinmacher, Gerosa, Sanchez, and Sarma}]{choudhuri2025needs}
Choudhuri R, Trinkenreich B, Pandita R, Kalliamvakou E, Steinmacher I, Gerosa M, Sanchez C, Sarma A (2025) What needs attention? prioritizing drivers of developers' trust and adoption of generative ai. arXiv preprint arXiv:250517418

\bibitem[{Chung et~al.(2012)Chung, Nixon, Yu, and Mylopoulos}]{chung2012non}
Chung L, Nixon BA, Yu E, Mylopoulos J (2012) Non-functional requirements in software engineering, vol~5. Springer Science \& Business Media

\bibitem[{Cronbach(1951)}]{cronbach1951coefficient}
Cronbach LJ (1951) Coefficient alpha and the internal structure of tests. psychometrika 16(3):297--334

\bibitem[{Davis(1989)}]{davis1989perceived}
Davis FD (1989) Perceived usefulness, perceived ease of use, and user acceptance of information technology. MIS quarterly pp 319--340

\bibitem[{Decan et~al.(2019)Decan, Mens, and Grosjean}]{decan2019empirical}
Decan A, Mens T, Grosjean P (2019) An empirical comparison of dependency network evolution in seven software packaging ecosystems. Empirical Software Engineering 24(1):381--416

\bibitem[{Ernst et~al.(2019)Ernst, Kazman, and Bianco}]{ernst2019component}
Ernst N, Kazman R, Bianco P (2019) Component comparison, evaluation, and selection: A continuous approach. In: 2019 IEEE International Conference on Software Architecture Companion (ICSA-C), IEEE, pp 87--90

\bibitem[{Farshidi(2020)}]{farshidi2020multi}
Farshidi S (2020) Multi-criteria decision-making in software production. PhD thesis, Utrecht University

\bibitem[{Farshidi(2025)}]{farshidi2025pypi}
Farshidi S (2025) {PyPI - Software Packages}. \url{https://doi.org/10.17632/n99pxfpf4x.1}, \doi{10.17632/n99pxfpf4x.1}

\bibitem[{Farshidi and Jansen(2020)}]{farshidi2020decision}
Farshidi S, Jansen S (2020) A decision support system for pattern-driven software architecture. In: European Conference on Software Architecture, Springer International Publishing Cham, pp 68--81

\bibitem[{Farshidi et~al.(2018{\natexlab{a}})Farshidi, Jansen, De~Jong, and Brinkkemper}]{Cloudfarshidi2018decision}
Farshidi S, Jansen S, De~Jong R, Brinkkemper S (2018{\natexlab{a}}) A decision support system for cloud service provider selection problem in software producing organizations. In: 2018 IEEE 20th Conference on Business Informatics (CBI), IEEE, vol~1, pp 139--148

\bibitem[{Farshidi et~al.(2018{\natexlab{b}})Farshidi, Jansen, De~Jong, and Brinkkemper}]{farshidi2018multiple}
Farshidi S, Jansen S, De~Jong R, Brinkkemper S (2018{\natexlab{b}}) Multiple criteria decision support in requirements negotiation. In: REFSQ Workshops

\bibitem[{Farshidi et~al.(2018{\natexlab{c}})Farshidi, Jansen, de~Jong, and Brinkkemper}]{Databasefarshidi2018decision}
Farshidi S, Jansen S, de~Jong R, Brinkkemper S (2018{\natexlab{c}}) A decision support system for software technology selection. Journal of Decision systems 27(sup1):98--110

\bibitem[{Farshidi et~al.(2020{\natexlab{a}})Farshidi, Jansen, Espa{\~n}a, and Verkleij}]{Blockchainfarshidi2020decision}
Farshidi S, Jansen S, Espa{\~n}a S, Verkleij J (2020{\natexlab{a}}) Decision support for blockchain platform selection: Three industry case studies. IEEE transactions on Engineering management 67(4):1109--1128

\bibitem[{Farshidi et~al.(2020{\natexlab{b}})Farshidi, Jansen, and van~der Werf}]{ArchitecturePatternsfarshidi2020capturing}
Farshidi S, Jansen S, van~der Werf JM (2020{\natexlab{b}}) Capturing software architecture knowledge for pattern-driven design. Journal of Systems and Software 169:110714

\bibitem[{Farshidi et~al.(2021{\natexlab{a}})Farshidi, Jansen, and Deldar}]{farshidi2021decision}
Farshidi S, Jansen S, Deldar M (2021{\natexlab{a}}) A decision model for programming language ecosystem selection: Seven industry case studies. Information and software technology 139:106640

\bibitem[{Farshidi et~al.(2021{\natexlab{b}})Farshidi, Jansen, and Fortuin}]{farshidi2021model}
Farshidi S, Jansen S, Fortuin S (2021{\natexlab{b}}) Model-driven development platform selection: four industry case studies. Software and Systems Modeling 20(5):1525--1551

\bibitem[{Farshidi et~al.(2024{\natexlab{a}})Farshidi, Kwantes, and Jansen}]{BPMLfarshidi2024business}
Farshidi S, Kwantes IB, Jansen S (2024{\natexlab{a}}) Business process modeling language selection for research modelers. Software and Systems Modeling 23(1):137--162

\bibitem[{Farshidi et~al.(2024{\natexlab{b}})Farshidi, Rezaee, Mazaheri, Rahimi, Dadashzadeh, Ziabakhsh, Eskandari, and Jansen}]{IntentModelfarshidi2024understanding}
Farshidi S, Rezaee K, Mazaheri S, Rahimi AH, Dadashzadeh A, Ziabakhsh M, Eskandari S, Jansen S (2024{\natexlab{b}}) Understanding user intent modeling for conversational recommender systems: a systematic literature review. User Modeling and User-Adapted Interaction pp 1--64

\bibitem[{Filgueira and Garijo(2022)}]{filgueira2022inspect4py}
Filgueira R, Garijo D (2022) Inspect4py: A knowledge extraction framework for python code repositories. In: Proceedings of the 19th International Conference on Mining Software Repositories, pp 232--236

\bibitem[{Fritz et~al.(2024)Fritz, Georg, Mele, and Schweinberger}]{fritz2024vulnerability}
Fritz C, Georg CP, Mele A, Schweinberger M (2024) Vulnerability webs: Systemic risk in software networks. arXiv preprint arXiv:240213375

\bibitem[{Garg et~al.(2021)Garg, Raheja, and Garg}]{garg2021decision}
Garg R, Raheja S, Garg RK (2021) Decision support system for optimal selection of software reliability growth models using a hybrid approach. IEEE Transactions on Reliability 71(1):149--161

\bibitem[{Gholamshahi and Hasheminejad(2019)}]{gholamshahi2019software}
Gholamshahi S, Hasheminejad SMH (2019) Software component identification and selection: A research review. Software: Practice and Experience 49(1):40--69

\bibitem[{{GitHub}(2023)}]{githubcopilot}
{GitHub} (2023) Github copilot: Your ai pair programmer. \url{https://github.com/features/copilot}, accessed July 28, 2025

\bibitem[{Grande et~al.(2014)Grande, Rodrigues, and Dias-Neto}]{grande2014framework}
Grande ADS, Rodrigues RDF, Dias-Neto AC (2014) A framework to support the selection of software technologies by search-based strategy. In: 2014 IEEE 26th International Conference on Tools with Artificial Intelligence, IEEE, pp 979--983

\bibitem[{Hamza et~al.(2025)Hamza, Farshidi, Akbar et~al.}]{Serverlesshamza2025decision}
Hamza M, Farshidi S, Akbar MA, et~al. (2025) A decision model for selecting serverless platforms. \url{https://doi.org/10.21203/rs.3.rs-5795769/v1}, pREPRINT (Version 1) available at Research Square

\bibitem[{Hevner et~al.(2004)Hevner, March, Park, and Ram}]{hevner2004design}
Hevner AR, March ST, Park J, Ram S (2004) Design science in information systems research. MIS quarterly pp 75--105

\bibitem[{Hou et~al.(2024)Hou, Feng, Farshidi, and Jansen}]{hou2024evaluating}
Hou F, Feng L, Farshidi S, Jansen S (2024) Evaluating software quality through user reviews: The isoftsentiment tool. In: International Conference on Product-Focused Software Process Improvement, Springer, pp 75--91

\bibitem[{{International Organization for Standardization}(2011)}]{iso25010}
{International Organization for Standardization} (2011) {ISO/IEC 25010:2011} systems and software engineering -- systems and software quality requirements and evaluation (square) -- system and software quality models. \url{https://iso.org/standard/35733.html}, accessed: 2025-07-30

\bibitem[{Ishio et~al.(2016)Ishio, Kula, Kanda, German, and Inoue}]{ishio2016software}
Ishio T, Kula RG, Kanda T, German DM, Inoue K (2016) Software ingredients: Detection of third-party component reuse in java software release. In: Proceedings of the 13th International Conference on Mining Software Repositories, pp 339--350

\bibitem[{Islam et~al.(2023)Islam, Kula, Treude, Chinthanet, Ishio, and Matsumoto}]{islam2023empirical}
Islam S, Kula RG, Treude C, Chinthanet B, Ishio T, Matsumoto K (2023) An empirical study of package management issues via stack overflow. IEICE TRANSACTIONS on Information and Systems 106(2):138--147

\bibitem[{Jabbarpour et~al.(2021)Jabbarpour, Saghiri, and Sookhak}]{jabbarpour2021framework}
Jabbarpour MR, Saghiri AM, Sookhak M (2021) A framework for component selection considering dark sides of artificial intelligence: a case study on autonomous vehicle. Electronics 10(4):384

\bibitem[{Jabborov et~al.(2023)Jabborov, Kharlamova, Kholmatova, Kruglov, Kruglov, and Succi}]{jabborov2023taxonomy}
Jabborov A, Kharlamova A, Kholmatova Z, Kruglov A, Kruglov V, Succi G (2023) Taxonomy of quality assessment for intelligent software systems: A systematic literature review. IEEE Access 11:130491--130507

\bibitem[{Jadhav and Sonar(2009)}]{jadhav2009evaluating}
Jadhav AS, Sonar RM (2009) Evaluating and selecting software packages: A review. Information and software technology 51(3):555--563

\bibitem[{Jadhav and Sonar(2011)}]{jadhav2011framework}
Jadhav AS, Sonar RM (2011) Framework for evaluation and selection of the software packages: A hybrid knowledge based system approach. Journal of Systems and Software 84(8):1394--1407

\bibitem[{Jha et~al.(2014)Jha, Bali, Narula, and Kalra}]{jha2014optimal}
Jha P, Bali V, Narula S, Kalra M (2014) Optimal component selection based on cohesion and coupling for component-based software system. In: Proceedings of the Second International Conference on Soft Computing for Problem Solving (SocProS 2012), December 28-30, 2012, Springer, pp 1499--1512

\bibitem[{Kalantari et~al.(2021)Kalantari, Motameni, Akbari, and Rabbani}]{kalantari2021optimal}
Kalantari S, Motameni H, Akbari E, Rabbani M (2021) Optimal components selection based on fuzzy-intra coupling density for component-based software systems under build-or-buy scheme. Complex \& Intelligent Systems 7(6):3111--3134

\bibitem[{Kaur et~al.(2015)Kaur, Arora, Jha, and Madan}]{kaurfuzzy}
Kaur R, Arora S, Jha P, Madan S (2015) Fuzzy multi-criteria approach for component selection of fault tolerant software system under consensus recovery block scheme. Procedia Computer Science 45:842--851

\bibitem[{Khandoker et~al.(2022)Khandoker, Sint, Gessl, Zeman, Jungreitmayr, Wahl, Wenigwieser, and Kretschmer}]{khandoker2022towards}
Khandoker A, Sint S, Gessl G, Zeman K, Jungreitmayr F, Wahl H, Wenigwieser A, Kretschmer R (2022) Towards a logical framework for ideal mbse tool selection based on discipline specific requirements. Journal of Systems and Software 189:111306

\bibitem[{Kiv et~al.(2018)Kiv, Heng, Kolp, and Wautelet}]{kiv2018agile}
Kiv S, Heng S, Kolp M, Wautelet Y (2018) Agile manifesto and practices selection for tailoring software development: A systematic literature review. In: Product-Focused Software Process Improvement: 19th International Conference, PROFES 2018, Wolfsburg, Germany, November 28--30, 2018, Proceedings 19, Springer, pp 12--30

\bibitem[{Kula et~al.(2017)Kula, Ouni, German, and Inoue}]{kula2017impact}
Kula RG, Ouni A, German DM, Inoue K (2017) On the impact of micro-packages: An empirical study of the npm javascript ecosystem. arXiv preprint arXiv:170904638

\bibitem[{Kulkarni and Rodd(2020)}]{kulkarni2020context}
Kulkarni S, Rodd SF (2020) Context aware recommendation systems: A review of the state of the art techniques. Computer Science Review 37:100255

\bibitem[{Kurilovas(2020)}]{kurilovas2020data}
Kurilovas E (2020) On data-driven decision-making for quality education. Computers in Human Behavior 107:105774

\bibitem[{Li et~al.(2024)Li, Cheng, Chen, Xuan, He, and Shang}]{li2024assessing}
Li S, Cheng Y, Chen J, Xuan J, He S, Shang W (2024) Assessing the performance of ai-generated code: A case study on github copilot. In: 2024 IEEE 35th International Symposium on Software Reliability Engineering (ISSRE), IEEE, pp 216--227

\bibitem[{Li et~al.(2016)Li, Wang, Wang, Yan, Xie, and Mei}]{li2016relationship}
Li X, Wang Z, Wang Q, Yan S, Xie T, Mei H (2016) Relationship-aware code search for javascript frameworks. In: Proceedings of the 2016 24th ACM SIGSOFT International Symposium on Foundations of Software Engineering, pp 690--701

\bibitem[{Lv et~al.(2015)Lv, Zhang, Lou, Wang, Zhang, and Zhao}]{lv2015codehow}
Lv F, Zhang H, Lou Jg, Wang S, Zhang D, Zhao J (2015) Codehow: Effective code search based on api understanding and extended boolean model (e). In: 2015 30th IEEE/ACM International Conference on Automated Software Engineering (ASE), IEEE, pp 260--270

\bibitem[{McNemar(1947)}]{mcnemar1947note}
McNemar Q (1947) Note on the sampling error of the difference between correlated proportions or percentages. Psychometrika 12(2):153--157

\bibitem[{Mehlawat et~al.(2020)Mehlawat, Gupta, and Mahajan}]{mehlawat2020multi}
Mehlawat MK, Gupta P, Mahajan D (2020) A multi-period multi-objective optimization framework for software enhancement and component evaluation, selection and integration. Information Sciences 523:91--110

\bibitem[{Mehta et~al.(2022)Mehta, Tandon, and Sharma}]{mehta2022integration}
Mehta P, Tandon A, Sharma H (2022) Integration of fahp and copras-g for software component selection. Optimization Models in Software Reliability pp 263--282

\bibitem[{Muhammed et~al.(2020)Muhammed, Mehmood, Abozinadah, and Sharaf}]{muhammed2020selecweb}
Muhammed T, Mehmood R, Abozinadah E, Sharaf S (2020) Selecweb: a software tool for automatic selection of web frameworks. Smart Infrastructure and Applications: Foundations for Smarter Cities and Societies pp 329--346

\bibitem[{Nabot et~al.(2025)Nabot, Al-Qerem, Omar, and Jebreen}]{nabot2025utilizing}
Nabot A, Al-Qerem A, Omar F, Jebreen I (2025) Utilizing evidential reasoning (er) approach for software components selection. SN Computer Science 6(3):220

\bibitem[{Nuh et~al.(2021)Nuh, Koh, Baharom, Osman, and Kew}]{nuh2021performance}
Nuh JA, Koh TW, Baharom S, Osman MH, Kew SN (2021) Performance evaluation metrics for multi-objective evolutionary algorithms in search-based software engineering: Systematic literature review. Applied Sciences 11(7):3117

\bibitem[{OpenAI(2023{\natexlab{a}})}]{openai2023chatgpt}
OpenAI (2023{\natexlab{a}}) Chatgpt: Optimizing language models for dialogue. \url{https://openai.com/chatgpt}, accessed July 28, 2025

\bibitem[{OpenAI(2023{\natexlab{b}})}]{openai2023gpt4}
OpenAI (2023{\natexlab{b}}) Gpt-4 technical report. \url{https://arxiv.org/abs/2303.08774}, arXiv:2303.08774

\bibitem[{Ouni et~al.(2017)Ouni, Kula, Kessentini, Ishio, German, and Inoue}]{ouni2017search}
Ouni A, Kula RG, Kessentini M, Ishio T, German DM, Inoue K (2017) Search-based software library recommendation using multi-objective optimization. Information and Software Technology 83:55--75

\bibitem[{Pace et~al.(2023)Pace, Tommasel, and Vazquez}]{pace2023javascript}
Pace AD, Tommasel A, Vazquez HC (2023) The javascript package selection task: A comparative experiment using an llm-based approach. CLEI Electronic Journal 27(2):4--1

\bibitem[{Padhy et~al.(2019)Padhy, Panigrahi, and Satapathy}]{padhy2019identifying}
Padhy N, Panigrahi R, Satapathy SC (2019) Identifying the reusable components from component-based system: proposed metrics and model. In: Information Systems Design and Intelligent Applications: Proceedings of Fifth International Conference INDIA 2018 Volume 2, Springer, pp 89--99

\bibitem[{Pandey et~al.(2024)Pandey, Singh, Wei, and Shankar}]{pandey2024transforming}
Pandey R, Singh P, Wei R, Shankar S (2024) Transforming software development: Evaluating the efficiency and challenges of github copilot in real-world projects. arXiv preprint arXiv:240617910

\bibitem[{Pearce et~al.(2025)Pearce, Ahmad, Tan, Dolan-Gavitt, and Karri}]{pearce2025asleep}
Pearce H, Ahmad B, Tan B, Dolan-Gavitt B, Karri R (2025) Asleep at the keyboard? assessing the security of github copilot’s code contributions. Communications of the ACM 68(2):96--105

\bibitem[{Peffers et~al.(2007)Peffers, Tuunanen, Rothenberger, and Chatterjee}]{peffers2007design}
Peffers K, Tuunanen T, Rothenberger MA, Chatterjee S (2007) A design science research methodology for information systems research. Journal of management information systems 24(3):45--77

\bibitem[{Penrose(1946)}]{penrose1946elementary}
Penrose LS (1946) The elementary statistics of majority voting. Journal of the Royal Statistical Society 109(1):53--57

\bibitem[{Petersen et~al.(2017)Petersen, Badampudi, Shah, Wnuk, Gorschek, Papatheocharous, Axelsson, Sentilles, Crnkovic, and Cicchetti}]{petersen2017choosing}
Petersen K, Badampudi D, Shah SMA, Wnuk K, Gorschek T, Papatheocharous E, Axelsson J, Sentilles S, Crnkovic I, Cicchetti A (2017) Choosing component origins for software intensive systems: In-house, cots, oss or outsourcing?—a case survey. IEEE Transactions on Software Engineering 44(3):237--261

\bibitem[{Pressman(2010)}]{pressman2001software}
Pressman RS (2010) Software Engineering: A Practitioner’s Approach, 5th edn. McGraw-Hill, New York

\bibitem[{Qi et~al.(2020{\natexlab{a}})Qi, He, Chen, Zhang, Dou, and Ni}]{qi2020data}
Qi L, He Q, Chen F, Zhang X, Dou W, Ni Q (2020{\natexlab{a}}) Data-driven web apis recommendation for building web applications. IEEE transactions on big data 8(3):685--698

\bibitem[{Qi et~al.(2020{\natexlab{b}})Qi, Zhang, Li, Wan, Wen, and Gong}]{qi2020spatial}
Qi L, Zhang X, Li S, Wan S, Wen Y, Gong W (2020{\natexlab{b}}) Spatial-temporal data-driven service recommendation with privacy-preservation. Information Sciences 515:91--102

\bibitem[{Rahkema et~al.(2024)Rahkema, Pfahl, and Ramler}]{rahkema2024impact}
Rahkema K, Pfahl D, Ramler R (2024) The impact of new package managers on the library dependency ecosystem. PeerJ Computer Science 10:e2617

\bibitem[{Ralph et~al.(2020)Ralph, Ali, Baltes, Bianculli, Diaz, Dittrich, Ernst, Felderer, Feldt, Filieri et~al.}]{ralph2020empirical}
Ralph P, Ali Nb, Baltes S, Bianculli D, Diaz J, Dittrich Y, Ernst N, Felderer M, Feldt R, Filieri A, et~al. (2020) Empirical standards for software engineering research. arXiv preprint arXiv:201003525

\bibitem[{Rodas-Silva et~al.(2019)Rodas-Silva, Galindo, Garc{\'\i}a-Guti{\'e}rrez, and Benavides}]{rodas2019resdec}
Rodas-Silva J, Galindo JA, Garc{\'\i}a-Guti{\'e}rrez J, Benavides D (2019) Resdec: online management tool for implementation components selection in software product lines using recommender systems. In: Proceedings of the 23rd International Systems and Software Product Line Conference-Volume B, pp 33--36

\bibitem[{Ruohonen et~al.(2021)Ruohonen, Hjerppe, and Rindell}]{ruohonen2021large}
Ruohonen J, Hjerppe K, Rindell K (2021) A large-scale security-oriented static analysis of python packages in pypi. In: 2021 18th International Conference on Privacy, Security and Trust (PST), IEEE, pp 1--10

\bibitem[{Sabbagh(2024)}]{sabbagh2024vulnerability}
Sabbagh Z (2024) Vulnerability landscapes and the impact of dependencies: Automated collection of security-related data for open-source software and their dependencies. Master's thesis, KTH Royal Institute of Technology

\bibitem[{Sabouri et~al.(2025)Sabouri, Eibl, Zhou, Ziyadi, Medvidovic, Lindemann, and Chattopadhyay}]{sabouri2025trust}
Sabouri S, Eibl P, Zhou X, Ziyadi M, Medvidovic N, Lindemann L, Chattopadhyay S (2025) Trust dynamics in ai-assisted development: Definitions, factors, and implications. \urlprefix\url{https://www.amazon.science/publications/trust-dynamics-in-ai-assisted-development-definitions-factors-and-implications}

\bibitem[{Sarrab and Rehman(2014)}]{sarrab2014empirical}
Sarrab M, Rehman OMH (2014) Empirical study of open source software selection for adoption, based on software quality characteristics. Advances in Engineering Software 69:1--11

\bibitem[{Schmidt et~al.(2013)Schmidt, Stal, Rohnert, and Buschmann}]{schmidt2013pattern}
Schmidt DC, Stal M, Rohnert H, Buschmann F (2013) Pattern-oriented software architecture, patterns for concurrent and networked objects. John Wiley \& Sons

\bibitem[{Sculley et~al.(2015)Sculley, Holt, Golovin, Davydov, Phillips, Ebner, Chaudhary, Young, Crespo, and Dennison}]{sculley2015hidden}
Sculley D, Holt G, Golovin D, Davydov E, Phillips T, Ebner D, Chaudhary V, Young M, Crespo JF, Dennison D (2015) Hidden technical debt in machine learning systems. Advances in neural information processing systems 28

\bibitem[{Sheng et~al.(2008)Sheng, Provost, and Ipeirotis}]{sheng2008get}
Sheng VS, Provost F, Ipeirotis PG (2008) Get another label? improving data quality and data mining using multiple, noisy labelers. In: Proceedings of the 14th ACM SIGKDD international conference on Knowledge discovery and data mining, pp 614--622

\bibitem[{Spracklen et~al.(2024)Spracklen, Wijewickrama, Sakib, Maiti, Viswanath, and Jadliwala}]{spracklen2024we}
Spracklen J, Wijewickrama R, Sakib A, Maiti A, Viswanath B, Jadliwala M (2024) We have a package for you! a comprehensive analysis of package hallucinations by code generating llms. arXiv preprint arXiv:240610279

\bibitem[{Spracklen et~al.(2025)Spracklen, Wijewickrama, Sakib, Maiti, Viswanath, and Jadliwala}]{spracklen2025package}
Spracklen J, Wijewickrama R, Sakib AHMN, Maiti A, Viswanath B, Jadliwala M (2025) Package hallucinations: How llms can invent vulnerabilities. Proceedings of the USENIX Security Symposium

\bibitem[{Stats(2025)}]{pypistats2025}
Stats P (2025) Pypi download statistics and project counts. \url{https://pypistats.org/packages/__all__}, accessed July 28, 2025

\bibitem[{Thakkar(2021)}]{thakkar2021multi}
Thakkar JJ (2021) Multi-criteria decision making, vol 336. Springer

\bibitem[{Thapar and Sarangal(2020)}]{thapar2020quantifying}
Thapar SS, Sarangal H (2020) Quantifying reusability of software components using hybrid fuzzy analytical hierarchy process (fahp)-metrics approach. Applied Soft Computing 88:105997

\bibitem[{{The Institute of Electrical and Electronics Engineers (IEEE)}(2025)}]{ieee2025taxonomy}
{The Institute of Electrical and Electronics Engineers (IEEE)} (2025) {IEEE} taxonomy for 2025. Version 2025 – IEEE Taxonomy; available from IEEE Standards Association

\bibitem[{Tidjon et~al.(2022)Tidjon, Rombaut, Khomh, Hassan et~al.}]{tidjon2022empirical}
Tidjon LN, Rombaut B, Khomh F, Hassan AE, et~al. (2022) An empirical study of library usage and dependency in deep learning frameworks. arXiv preprint arXiv:221115733

\bibitem[{Vazquez et~al.(2023)Vazquez, Diaz-Pace, Vidal, and Marcos}]{vazquez2023recommender}
Vazquez HC, Diaz-Pace JA, Vidal SA, Marcos C (2023) A recommender system for recovering relevant javascript packages from web repositories. In: 2023 IEEE 20th International Conference on Software Architecture (ICSA), IEEE, pp 175--185

\bibitem[{Wang(2023)}]{wang2023power}
Wang Y (2023) The power of openness-how open source software is reshaping software engineering and industrial adoption. Authorea Preprints

\bibitem[{Wiegers and Beatty(2013)}]{wiegers2013software}
Wiegers K, Beatty J (2013) Software requirements. Pearson Education

\bibitem[{Wilcoxon(1992)}]{wilcoxon1992individual}
Wilcoxon F (1992) Individual comparisons by ranking methods. In: Breakthroughs in statistics: Methodology and distribution, Springer, pp 196--202

\bibitem[{Wohlin(2014)}]{wohlin2014guidelines}
Wohlin C (2014) Guidelines for snowballing in systematic literature studies and a replication in software engineering. In: Proceedings of the 18th international conference on evaluation and assessment in software engineering, pp 1--10

\bibitem[{Wohlin et~al.(2012)Wohlin, Runeson, H{\"o}st, Ohlsson, Regnell, Wessl{\'e}n et~al.}]{wohlin2012experimentation}
Wohlin C, Runeson P, H{\"o}st M, Ohlsson MC, Regnell B, Wessl{\'e}n A, et~al. (2012) Experimentation in software engineering, vol 236. Springer

\bibitem[{Ying and Robillard(2014)}]{ying2014selection}
Ying AT, Robillard MP (2014) Selection and presentation practices for code example summarization. In: Proceedings of the 22nd acm sigsoft international symposium on foundations of software engineering, pp 460--471

\bibitem[{Zajdel et~al.(2022)Zajdel, Costa, and Mili}]{Zajdel2022OSS}
Zajdel S, Costa DE, Mili H (2022) Open source software: an approach to controlling usage and risk in application ecosystems. In: Proceedings of the 26th ACM International Systems and Software Product Line Conference - Volume A, Association for Computing Machinery, New York, NY, USA, SPLC '22, p 154–163, \doi{10.1145/3546932.3547000}, \urlprefix\url{https://doi.org/10.1145/3546932.3547000}

\end{thebibliography}

\nocite{pace2023javascript,
vazquez2023recommender,
hou2024evaluating,
jadhav2011framework,
nabot2025utilizing,
bibi2021conceptual,
jabbarpour2021framework,
kalantari2021optimal,
mehta2022integration,
rodas2019resdec,
ernst2019component,
mehlawat2020multi,
chatzipetrou2020component,
borg2019selecting,
petersen2017choosing,
gholamshahi2019software,
badampudi2016software,
sarrab2014empirical,
jha2014optimal,
thapar2020quantifying,
padhy2019identifying,
ouni2017search,
cervantes2019data,
ishio2016software,
li2016relationship,
ying2014selection,
lv2015codehow,
filgueira2022inspect4py,
grande2014framework,
kaurfuzzy,
farshidi2021decision,
qi2020data,
ArchitecturePatternsfarshidi2020capturing,
capilla2020teaching,
IntentModelfarshidi2024understanding,
abdulwareth2021toward,
ansari2020fuzzy,
kiv2018agile,
qi2020spatial,
muhammed2020selecweb,
khandoker2022towards,
farshidi2021model,
Oracleahmadjee2025decision,
Blockchainfarshidi2020decision,
DAObaninemeh2023decision,
BPMLfarshidi2024business,
akbar2023multi,
alshamsi2023multi,
garg2021decision,
nuh2021performance,
kurilovas2020data,
Cloudfarshidi2018decision,
Databasefarshidi2018decision}

\end{document}